\documentclass[numberedappendix]{emulateapj}
\usepackage{aas_macros}

\usepackage{color}
\usepackage[dvipsnames,svgnames]{xcolor}

\usepackage{natbib}

\usepackage{amsmath}
\usepackage{xspace}

\newcommand{\feh}         {\mbox{[Fe/H]}}
\newcommand{\kms}         {km~s$^{-1}$}
\newcommand{\masyr}       {mas~yr$^{-1}$}
\newcommand{\mua}         {\mu_{\alpha}\cos{\delta}}
\newcommand{\mud}         {\mu_{\delta}}

\def\spose#1{\hbox to 0pt{#1\hss}}
\def\lta{\mathrel{\spose{\lower 3pt\hbox{$\mathchar"218$}}
     \raise 2.0pt\hbox{$\mathchar"13C$}}}
\def\gta{\mathrel{\spose{\lower 3pt\hbox{$\mathchar"218$}}
    \raise 2.0pt\hbox{$\mathchar"13E$}}}

\shorttitle{\emph{Gaia} Proper Motions of Ultra-Faint Satellites}
\shortauthors{J. D. Simon}

\begin{document}

\title{\emph{Gaia} Proper Motions and Orbits of the Ultra-Faint Milky
  Way Satellites}

\author{Joshua D. Simon}

\affil{Observatories of the Carnegie Institution for
  Science, 813 Santa Barbara St., Pasadena, CA 91101}
\email{jsimon@carnegiescience.edu}

\begin{abstract}
The second data release from the \emph{Gaia} mission (DR2) provides a
comprehensive and unprecedented picture of the motions of astronomical
sources in the plane of the sky, extending from the solar neighborhood
to the outer reaches of the Milky Way.  I present proper motion
measurements based on \emph{Gaia} DR2 for 17 ultra-faint dwarf
galaxies within 100~kpc of the Milky Way.  I compile the
spectroscopically-confirmed member stars in each dwarf bright enough
for \emph{Gaia} astrometry from the literature, producing member
samples ranging from 2 stars in Triangulum~II to 68 stars in
Bo{\"o}tes~I.  From the spectroscopic member catalogs I estimate the
proper motion of each system.  I find good agreement with the proper
motions derived by the \emph{Gaia} collaboration for Bo{\"o}tes~I and
Leo~I.  The tangential velocities for 14 of the 17 dwarfs are
determined to better than 50~\kms, more than doubling the sample of
such measurements for Milky Way satellite galaxies.  The orbital
pericenters are well-constrained, with a median value of 38~kpc.  Only
one satellite, Tucana~III, is on an orbit passing within 15~kpc of the
Galactic center, suggesting that the remaining ultra-faint dwarfs are
unlikely to have experienced severe tidal stripping.  As a group, the
ultra-faint dwarfs are on high-velocity, eccentric, retrograde
trajectories, with nearly all of them having space motions exceeding
370~\kms.  A large majority of the objects are currently close to the
pericenters of their orbits.  In a low-mass (M$_{\rm vir} = 0.8 \times
10^{12}$~M$_{\odot}$) Milky Way potential, eight out of the 17
galaxies lack well-defined apocenters and appear likely to be on their
first infall, indicating that the Milky Way mass may be larger than
previously estimated or that many of the ultra-faint dwarfs are
associated with the Magellanic Clouds.  The median eccentricity of the
ultra-faint dwarf orbits is 0.79, similar to the values seen in
numerical simulations, but distinct from the rounder orbits of the
more luminous dwarf spheroidals.
\end{abstract}


\section{INTRODUCTION}
\label{intro}

The orbits of dwarf galaxies around the Milky Way can provide crucial
information regarding the mass and mass profile of the Galaxy
\citep[e.g.,][]{boylan-kolchin13}, the formation and evolution of
dwarf galaxies \citep[e.g.,][]{penarrubia08}, the history of the Local
Group \citep[e.g.,][]{rocha12}, and perhaps even the properties of
dark matter \citep[e.g.,][]{mw10}.  A particularly striking example is
the Magellanic Clouds and the origin of the Magellanic Stream: the
large tangential velocity of the Clouds measured by
\citet{kallivayalil06,kallivayalil13} suggests that the LMC and SMC
have completed at most one passage around the Milky Way
\citep{besla07}, which significantly changes models for the formation
of the stream \citep{nidever08,besla10,db11,db12,df16}.

Until now, proper motions of Milky Way satellite galaxies have been
determined painstakingly, either with multiple epochs of \emph{Hubble
  Space Telescope} imaging
\citep[e.g.,][]{piatek02,piatek03,lepine11,sohn13,pryor15} or with
longer time baselines from ground-based data
\citep[e.g.,][]{dinescu05,mendez10,dinescu16,dinescu18}.  These
measurements have been limited to the Magellanic Clouds and the
relatively luminous ($L \gtrsim 2\times10^{5}$~L$_{\odot}$) classical
dwarf spheroidal (dSph) galaxies and have typical uncertainties of
$\sim0.1$~mas~yr$^{-1}$.  In some cases, measurements by different
groups and with different techniques disagree by more than the quoted
uncertainties, suggesting the possibility of systematic errors
\citep{piatek07,mendez10,pryor15,dinescu16}.  The only existing proper
motions for ultra-faint ($L \lesssim 10^{5}$~L$_{\odot}$) dwarf
galaxies are the recent determination for Segue~1 by \citet{fritz17}
and the \emph{Gaia} team's just-released measurement of Bo{\"o}tes~I
\citep{gaiadr2helmi}.

With the second data release \citep[DR2;][]{gaiadr2brown} from the
\emph{Gaia} mission \citep{gaia16a}, proper motion measurements for Milky Way
satellites will become routine.  For objects nearby enough to have
appreciable motions in the plane of the sky, identification of bright
member stars in a satellite will be sufficient to determine its
tangential velocity with accuracy similar to the best presently
existing measurements.  The relationship between proper motion and
transverse velocity is

\begin{equation}
v_{tan} = 474~\left(\frac{\mu}{1~{\rm mas~yr}^{-1}}\right)~\left(\frac{d}{100~{\rm kpc}}\right)~{\rm km~s}^{-1},
\end{equation}

\noindent
where $\mu$ is the proper motion in \masyr\ and $d$ is the distance in
kpc.  Thus, a proper motion accuracy of 0.1~\masyr\ corresponds to a
velocity accuracy better than 50~\kms\ at 100~kpc, comparable to the
\emph{HST} proper motions of the classical dSphs.  In the \emph{Gaia} DR2
catalog, the proper motion for a single star at magnitude $\sim16$,
such as a giant near the tip of the red giant branch at $\sim60$~kpc,
is measured at this level.  For a dwarf galaxy containing multiple
stars at magnitude $\sim18$, more accurate measurements can be
obtained by averaging the \emph{Gaia} proper motions for all of the known
member stars.  

To date, proper motion measurements, and hence three-dimensional
velocity vectors, have only been published for 13 Milky Way
satellites.  A substantially larger set of space motions in the outer
halo of the Milky Way can now be determined.  In concert with new
constraints from stellar streams \citep{bh18}, these measurements will
provide a significant new sample of tracers of the Milky Way's
gravitational potential for inferring the mass of the Galaxy
\citep[e.g.,][]{battaglia05,sales07,lux10,watkins10,boylan-kolchin13,barber14,patel18}.

In this paper I present the first proper motion measurements for a
large sample of nearby ultra-faint dwarf galaxies.  In
Section~\ref{sec:members} I compile confirmed member stars in each
galaxy that are bright enough for accurate \emph{Gaia} astrometry.  In
Section~\ref{sec:validation} I test for possible systematics in the
\emph{Gaia} proper motions of distant and faint stars using systems with
well-measured proper motions.  In Section~\ref{sec:pms} I use the
member catalogs for each dwarf to determine their proper motions.  I
then compute the corresponding orbits around the Milky Way in
Section~\ref{sec:orbits}.  I discuss the implications of the orbits in
Section~\ref{sec:discussion} and conclude in
Section~\ref{sec:conclusions}.

\section{MEMBER SAMPLES}
\label{sec:members}

Motivated by the lack of available proper motions for ultra-faint
dwarf galaxies and the velocity accuracy discussed above, for this
study I select a sample of dwarf galaxies with $M_{V} > -8$ and $d <
100$~kpc.  Although there are 25 confirmed or likely dwarfs meeting
these criteria, not all of those objects have published spectroscopy,
meaning that member stars cannot be confidently separated from
non-members.  Therefore, for the remainder of this paper I limit
consideration to the 17 ultra-faint dwarfs for which spectroscopic
member catalogs are available.  When possible, I supplement the
spectroscopic samples with RR~Lyrae stars that have been identified
from their light curves.

I begin by compiling from the literature all known member stars, as
well as the confirmed non-members, in each of these 17 systems.  I
only consider stars brighter than $g=20$ because that is approximately
where the \emph{Gaia} DR2 proper motion accuracy reaches $\sim1$~\masyr.  I
cross-match the spectroscopic catalogs against public photometric data
from the Sloan Digital Sky Survey \citep[SDSS;][]{york00} for dwarfs
discovered in SDSS imaging, Pan-STARRS \citep{chambers16} for dwarfs
discovered in Pan-STARRS imaging, and the Dark Energy Survey
\citep[DES;][]{desdr1} for dwarfs discovered in DES imaging.  The most
recently-found dwarfs (Carina~II, Carina~III, and Hydrus~I) were
identified in imaging surveys near the south celestial pole that are
not currently public, so for those objects I use the coordinates and
magnitudes provided in the discovery papers.  Although the filters and
photometric systems for SDSS, Pan-STARRS, and DECam are not identical,
I do not attempt to place the photometry on a common system, so the
magnitudes reported here are heterogeneous at the level of a few
percent.

In the Appendix I provide a detailed discussion of the sources of the
spectroscopy and the size of the available member/non-member samples
for each dwarf, in order of their right ascension.  The assumed
positions, distances, and velocities for the target galaxies are
listed in Table~\ref{litdata_table}.

\begin{deluxetable*}{lccccrrr}
\tablecaption{Ultra-Faint Dwarf Positions, Distances, and Velocities}
\tablewidth{0pt}
\tablehead{
\colhead{Dwarf} &
\colhead{R.A. (J2000)} &
\colhead{Dec. (J2000)} & 
\colhead{Distance} &
\colhead{$v_{\rm hel}$} &
\colhead{Position} &
\colhead{Distance} &
\colhead{Velocity} \\
\colhead{} &
\colhead{(deg)} &
\colhead{(deg)} &
\colhead{(kpc)} &
\colhead{(\kms)} &
\colhead{reference} &
\colhead{reference} &
\colhead{reference} 
}
\startdata
 Triangulum~II &  33.3225 & \phs$36.1783$ & $ 30.0 \pm  2.0$ &    $- 381.7 \pm  1.1$ &  1 &  1 &  2 \\
       Segue~2 &  34.8167 & \phs$20.1753$ & $ 35.0 \pm  2.0$ &    $-  40.2 \pm  0.9$ &  3 &  3 &  4 \\
      Hydrus~I &  37.3890 & $   -79.3089$ & $ 27.6 \pm  0.5$ & \phs$  80.4 \pm  0.6$ &  5 &  5 &  5 \\
  Horologium~I &  43.8700 & $   -54.1100$ & $ 87.0 \pm 12.0$ & \phs$ 112.8 \pm  2.6$ &  6 &  6 &  7 \\
  Reticulum~II &  53.9200 & $   -54.0500$ & $ 32.0 \pm  3.0$ & \phs$  62.8 \pm  0.5$ &  6 &  6 &  8 \\
     Carina~II & 114.1066 & $   -57.9991$ & $ 36.2 \pm  0.6$ & \phs$ 477.2 \pm  1.2$ &  9 &  9 & 10 \\
    Carina~III & 114.6298 & $   -57.8997$ & $ 27.8 \pm  0.6$ & \phs$ 284.6 \pm  3.4$ &  9 &  9 & 10 \\
 Ursa~Major~II & 132.8744 & \phs$63.1331$ & $ 34.7 \pm  2.0$ &    $- 116.5 \pm  1.9$ & 11 & 12 & 13 \\
       Segue~1 & 151.7633 & \phs$16.0736$ & $ 23.0 \pm  2.0$ & \phs$ 208.5 \pm  0.9$ & 14 & 15 & 16 \\
  Ursa~Major~I & 158.6850 & \phs$51.9261$ & $ 97.3 \pm  6.0$ &    $-  55.3 \pm  1.4$ & 17 & 18 & 13 \\
     Willman~1 & 162.3413 & \phs$51.0528$ & $ 45.0 \pm 10.0$ &    $-  12.8 \pm  1.0$ & 14 & 19 & 20 \\
Coma~Berenices & 186.7458 & \phs$23.9076$ & $ 42.0 \pm  2.0$ & \phs$  98.1 \pm  0.9$ & 11 & 21 & 13 \\
     Bo{\"o}tes~II & 209.5213 & \phs$12.8586$ & $ 42.0 \pm  2.0$ &    $- 117.0 \pm  5.2$ & 22 & 22 & 23 \\
      Bo{\"o}tes~I & 210.0225 & \phs$14.5006$ & $ 66.0 \pm  2.0$ & \phs$ 101.8 \pm  0.7$ & 24 & 25 & 26 \\
      Draco~II & 238.1983 & \phs$64.5653$ & $ 20.0 \pm  3.0$ &    $- 347.6 \pm  1.8$ & 27 & 27 & 28 \\
     Tucana~II & 343.0600 & $   -58.5700$ & $ 58.0 \pm  8.0$ &    $- 129.1 \pm  3.5$ &  6 &  6 & 29 \\
    Tucana~III & 359.1500 & $   -59.6000$ & $ 25.0 \pm  2.0$ &    $- 102.3 \pm  0.4$ & 30 & 30 & 31 \\
\enddata
\tablecomments{References:  (1)  \citet{laevens15};  (2)  \citet{kirby17};  (3)  \citet{belokurov09};  (4)  \citet{kirby13};  (5)  \citet{koposov18};  (6)  \citet{bechtol15};  (7)  \citet{koposov15};  (8)  \citet{simon15};  (9)  \citet{torrealba18};  (10)  \citet{li18};  (11)  \citet{munoz10};  (12)  \citet{dallora12};  (13)  \citet{sg07};  (14)  \citet{martin08};  (15)  \citet{belokurov07};  (16)  \citet{simon11};  (17)  \citet{okamoto08};  (18)  \citet{garofalo13};  (19)  \citet{willman05a};  (20)  \citet{willman11};  (21)  \citet{musella09};  (22)  \citet{walsh08};  (23)  \citet{koch09};  (24)  \citet{okamoto12};  (25)  \citet{dallora06};  (26)  \citet{koposov11};  (27)  \citet{laevens15b};  (28)  \citet{martin16b};  (29)  \citet{walker16};  (30)  \citet{drlica15};  (31)  \citet{simon17}. }
\label{litdata_table}
\end{deluxetable*}

\section{VALIDATION}
\label{sec:validation}

Because the \emph{Gaia} DR2 data were just released, no independent checks of
the astrometric performance in the regime of faint and distant sources
are available yet.  In this section I carry out some simple tests to
assess the reliability of the dwarf galaxy proper motions I determine
in Section~\ref{sec:pms}.  \citet{gaiadr2lindegren} find using an
all-sky sample of quasars that on scales of $<1\degr$ there are
systematics in the DR2 proper motions with a root-mean-square value of
0.066~\masyr.  They suggest that averaging of many faint sources
cannot reduce the proper motion uncertainties below this floor.
\citet{gaiadr2helmi} use the largest dwarf galaxies on the sky, the
Magellanic Clouds and Sagittarius, to provide an independent
constraint on proper motion systematics and estimate an overall
minimum systematic uncertainty of $\sim0.035$~\masyr.

\subsection{$\Omega$ Centauri}

One of the most accurate extragalactic proper motions currently
available is that determined by \citet{libralato18} for
$\Omega$~Centauri.  I query the \emph{Gaia} DR2 catalog over a circle of
radius $1.5\arcmin$ area centered on the coordinates of the
\citeauthor{libralato18} field, approximately matching their
$2.73\arcmin \times 2.73\arcmin$ \emph{HST}
coverage.\footnote{Although this area does not exactly match that of
  the \emph{HST} pointing, given the large distance of this field from
  the center of $\Omega$~Cen, it is not important that an identical
  set of stars be used.}  As shown by \citet{libralato18}, the
foreground contamination in this region from Milky Way stars with
proper motions similar to that of $\Omega$~Cen is quite small.

The stars in this field with \emph{Gaia} proper motions broadly consistent
with $\Omega$~Cen membership (within 10~\masyr\ of that reported for
$\Omega$~Cen by \citealt{libralato18}) have median proper motions of
$\mua = -3.26$~\masyr\ and $\mud = -6.54$~\masyr.  I select those
stars with proper motions within 2~\masyr\ of these values as the
highest quality sample in this field.  From the remaining 217 stars I
determine a weighted average proper motion of $\mua = -3.295 \pm
0.025$~\masyr, $\mud = -6.615 \pm 0.033$~\masyr.  The \emph{HST}
proper motion in this field is $\mua = -3.341 \pm 0.028$~\masyr, $\mud
= -6.557 \pm 0.043$~\masyr.  The \emph{Gaia} and \emph{HST} measurements
agree at the $\sim1\sigma$ level; if I included the systematic
uncertainties listed above the agreement would be better than
$1\sigma$.

\subsection{Leo~I}
\label{sec:leo1test}

\begin{figure}[th!]
\epsscale{1.2}
\plotone{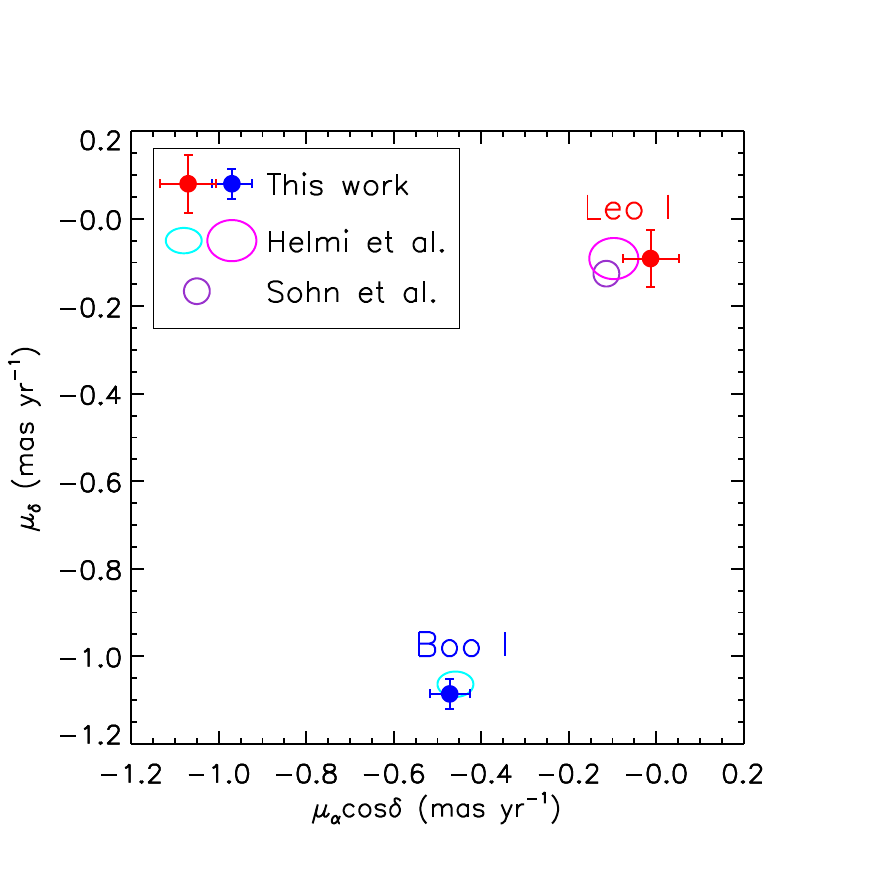}
\caption{Comparison of my proper motion measurements with those of
  \citet{gaiadr2helmi} for Leo~I and Boo~I.  The solid points (red
  for Leo~I and blue for Boo~I) illustrate the proper motions I find
  in Sections~\ref{sec:leo1test} and \ref{sec:boo1test}, while the
  magenta (Leo~I) and cyan (Boo~I) ellipses represent the results from
  the \emph{Gaia} collaboration.  The purple circle for Leo~I shows the
  \emph{HST} measurement by \citet{sohn13}.}
\label{litpm_comparison}
\end{figure}

The classical dSph with the most accurately measured proper motion is
Leo~I, with uncertainties of $\sim0.03$~\masyr\ per coordinate
\citep{sohn13}.  Leo~I is a difficult target for \emph{Gaia} because of its
large distance \citep[255~kpc;][]{held01,bellazzini04}, so that its
proper motion is small and its brightest stars are at $V \approx 19$
(fainter than those in most of the ultra-faint dwarfs).  On the other
hand, it contains many more stars whose proper motions can be averaged
together to reduce uncertainties.  I select 202 Leo~I members brighter
than $V=20$ from the spectroscopic catalog of \citet{kirby10}.  Of
these, 187 stars have proper motions in the DR2 catalog, and the
weighted average of this sample is $\mua = -0.013 \pm 0.064$~\masyr,
$\mud = -0.091 \pm 0.066$~\masyr.  For comparison, \citet{sohn13}
measure proper motions of $\mua = -0.1140 \pm 0.0295$~\masyr, $\mud =
-0.1256 \pm 0.0293$~\masyr.  Because the Leo~I stars are so faint the
\emph{Gaia} proper motion has larger uncertainties than those determined by
\emph{HST}.  Still, the proper motion in declination agrees at better
than $1\sigma$, and that in right ascension differs by $\sim1.4\sigma$
(again neglecting systematic uncertainties).  Using a photometric DR2
sample rather than my spectroscopic one, \citet{gaiadr2helmi} find
$\mua = -0.097 \pm 0.056$~\masyr, $\mud = -0.091 \pm 0.047$~\masyr, in
agreement with my measurement within the uncertainties (see
Figure~\ref{litpm_comparison}).

\subsection{Bo{\"o}tes~I}
\label{sec:boo1test}

My sample of ultra-faint dwarf galaxies has one object in common with
that of \citet{gaiadr2helmi}: Boo~I.  Using a
photometrically-selected sample as for Leo~I, \citet{gaiadr2helmi}
measure proper motions of $\mua = -0.459 \pm 0.041$~\masyr, $\mud =
-1.064 \pm 0.029$~\masyr\ from 115 stars.  After rejecting 5 stars
with deviant proper motions from the spectroscopic sample described in
the Appendix, I measure $\mua = -0.472 \pm 0.046$~\masyr, $\mud =
-1.086 \pm 0.034$~\masyr\ from 63 stars, in perfect agreement with
\citet{gaiadr2helmi} as shown in Fig.~\ref{litpm_comparison}.  The 5
astrometric non-members include: 3 stars at very large separations
($>26\arcmin$) from Boo~I from the sample of \citet{norris10}, one RRc
variable 20\arcmin\ from the center of Boo~I and with a similar mean
magnitude to the other Boo~I RR~Lyrae but a larger photometric
amplitude \citep{siegel06}, and one star from \citet{koposov11} within
the main body of Boo~I but with a radial velocity offset from the
systemic velocity of the galaxy by $\sim2\sigma$.

\subsection{Background Quasars}

Alternatively, instead of comparing the very small measured motions of
stars by \emph{Gaia} with the measured motions of similar stars from
independent (but challenging) \emph{HST} observations, one can check
very distant extragalactic sources that can be assumed to be fixed on
the sky.  Of course, the \emph{Gaia} team has already carried out
extensive tests along these lines \citep[e.g.,][]{gaiadr2lindegren}.
However, for completeness I perform this test in small areas around
the ultra-faint dwarfs in case there are local systematic errors in
any of these fields.  I select spectroscopically-confirmed quasars
from SDSS DR14 \citep{sdssdr14} within 30\arcmin\ of each of the
dwarfs within the SDSS footprint.  Across the fields of 7 ultra-faint
dwarfs (UMa~II, Segue~1, UMa~I, Willman~1, Com~Ber, Boo~II, and Boo~I)
I find a total of 154 SDSS quasars brighter than $g=20$ that have
counterparts in the \emph{Gaia} DR2 catalog.  The average proper
motions in each field are consistent with a population of sources with
zero net proper motion and Gaussian errors accurately described by the
DR2 uncertainties.  There is no evidence for a bias in \emph{Gaia}
proper motions near any of these dwarfs down to a level of
$\sim0.1$~\masyr.  However, since the number of quasars in each field
is small and the quasars are generally faint, this comparison is not
sensitive to errors of the size of the estimated DR2 systematic
uncertainties.


Based on the above comparisons with both an independent \emph{Gaia} DR2
analysis and with \emph{HST} data sets, I conclude that accurate DR2
proper motions can be straightforwardly measured for a number of
extragalactic systems, and that any systematic uncertainties are
comparable to or smaller than the statistical errors.

\section{PROPER MOTIONS}
\label{sec:pms}

\begin{figure*}[th!]
\epsscale{1.2}
\plotone{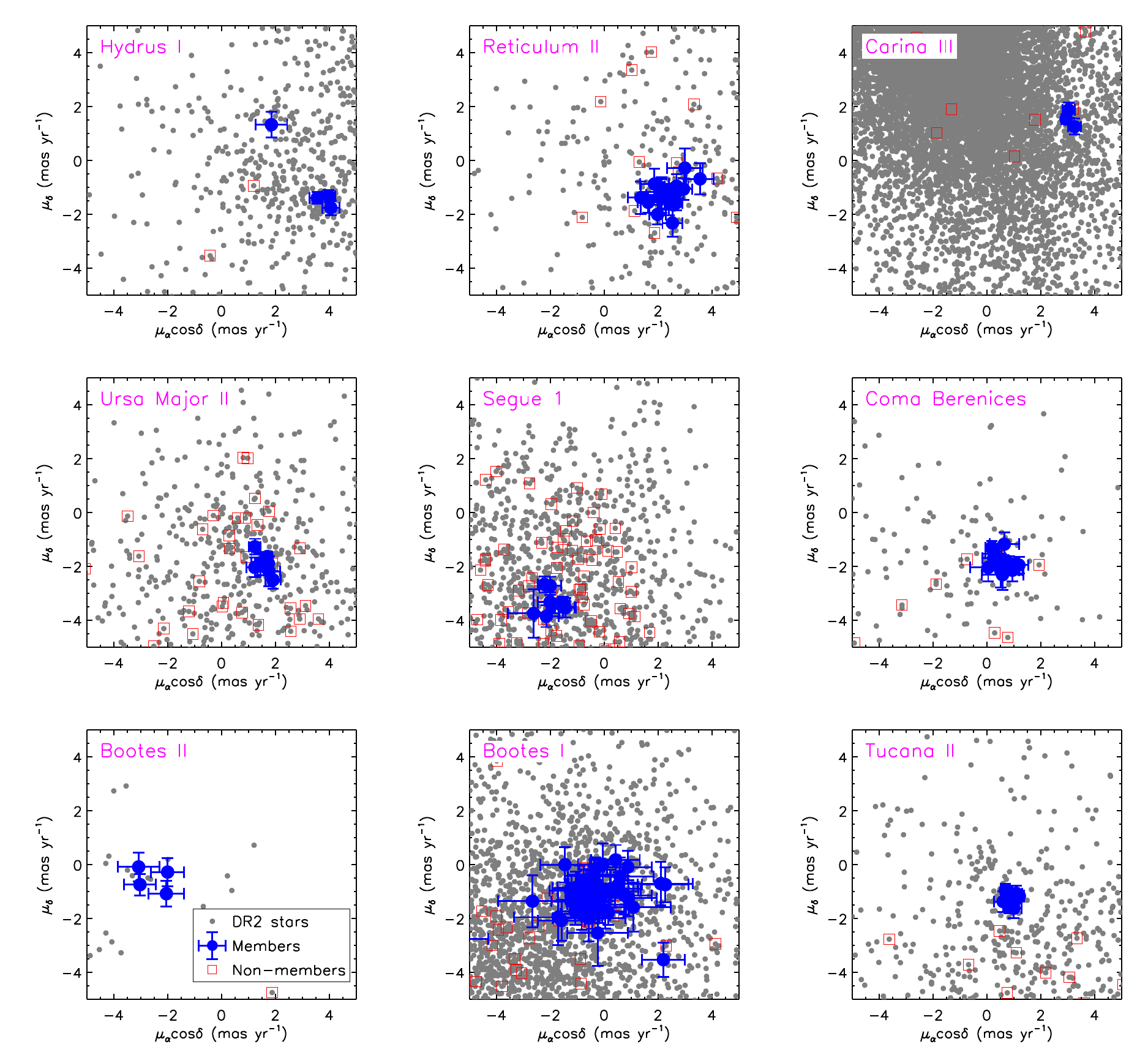}
\caption{\emph{Gaia} DR2 proper motion diagrams of a subset of the target
  galaxies.  The area over which stars are selected from the DR2
  catalog varies from galaxy to galaxy depending on the extent of the
  spectroscopic coverage; stars are selected out to a radius just
  beyond the most distant confirmed members/non-members.  All stars in
  the field are plotted as small gray dots, member stars are shown as
  blue circles, and spectroscopic non-members are plotted as red
  squares.  Even for objects with very small spectroscopic samples
  such as Hyi~I, Car~III, and Boo~II it is clear that the members
  exhibit coherent proper motions that are distinct from those of
  other stars in the field.}
\label{pmdiagrams}
\end{figure*}

For each of the ultra-faint dwarfs, I cross-match the confirmed member
lists assembled in Section~\ref{sec:members} with the \emph{Gaia} DR2
catalog.  I determine the proper motion of each system by taking a
weighted average of the DR2 proper motions of the member stars.
Because the uncertainties on these averages are already well above
than the systematic error floor determined by \citet{gaiadr2helmi} in
most cases, I do not apply any additional correction to the derived
uncertainties.  In general, it is obvious from inspecting the
distributions of proper motions for the known members, known
non-members, and field stars that the member samples are free from
contaminants and form tight distributions in proper motion space (see
Figure~\ref{pmdiagrams}).  A catalog of the stars used in this
analysis is provided in Table~\ref{mem_star_table}.

\tabletypesize{\scriptsize}
\begin{deluxetable*}{lcccccccll}
\tablecaption{\emph{Gaia} DR2 Proper Motions of Stars in Ultra-Faint Dwarf Galaxies}
\tablewidth{0pt}
\tablehead{
\colhead{ID} &
\colhead{Dwarf} &
\colhead{R.A.} &
\colhead{Dec.} & 
\colhead{g} &
\colhead{r} &
\colhead{$\mua$} &
\colhead{$\mud$} &
\colhead{References\tablenotemark{a}} &
\colhead{Comment} \\
\colhead{} &
\colhead{} &
\colhead{(deg)} &
\colhead{(deg)} &
\colhead{} &
\colhead{} &
\colhead{(\masyr)} &
\colhead{(\masyr)} &
\colhead{} &
\colhead{} 
}
\startdata
 PS1J02131655+3610458    & Tri II &  33.31894 &  36.17938 & 17.60 & 17.01 & $  0.406 \pm  0.209$ & $  0.511 \pm  0.176$ &                1,2,3,4 &              \\ 
 PS1J02132154+3609574    & Tri II &  33.33976 &  36.16594 & 19.29 & 18.81 & $  2.044 \pm  0.499$ & $  1.093 \pm  0.439$ &                1,2,3,4 &              \\ 
SDSSJ02191849+2010219   & Segue 2 &  34.82706 &  20.17275 & 18.76 & 18.81 & $  2.327 \pm  0.818$ & $  1.253 \pm  0.753$ &                    5,6 &              \\ 
SDSSJ02192430+2010168   & Segue 2 &  34.85125 &  20.17133 & 18.93 & 18.19 & $  1.060 \pm  0.520$ & $ -0.813 \pm  0.316$ &                    5,6 &              \\ 
SDSSJ02190759+2012209   & Segue 2 &  34.78161 &  20.20578 & 18.84 & 18.78 & $  0.805 \pm  0.926$ & $  0.085 \pm  0.498$ &                    5,6 &              \\ 
SDSSJ02190493+2007154   & Segue 2 &  34.77053 &  20.12095 & 19.98 & 19.24 & $  1.182 \pm  0.883$ & $  0.758 \pm  0.597$ &                    5,6 &              \\ 
SDSSJ02190004+2009457   & Segue 2 &  34.75018 &  20.16270 & 19.60 & 18.84 & $  0.452 \pm  0.622$ & $ -0.956 \pm  0.431$ &                    5,6 &              \\ 
SDSSJ02193468+2011443   & Segue 2 &  34.89448 &  20.19565 & 19.25 & 19.34 & $  2.007 \pm  1.254$ & $ -1.753 \pm  0.658$ &                    5,6 &              \\ 
SDSSJ02192271+2004433   & Segue 2 &  34.84463 &  20.07870 & 19.55 & 18.81 & $  2.163 \pm  0.626$ & $  0.233 \pm  0.508$ &                    5,6 &              \\ 
SDSSJ02195535+2007492   & Segue 2 &  34.98061 &  20.13034 & 19.18 & 18.49 & $  2.660 \pm  0.668$ & $  0.023 \pm  0.352$ &                    5,6 &              \\ 
SDSSJ02190006+2006351   & Segue 2 &  34.75026 &  20.10978 & 19.16 & 18.61 & $  1.309 \pm  0.521$ & $ -0.895 \pm  0.396$ &                    6,7 &              \\ 
SDSSJ02191711+2009306   & Segue 2 &  34.82127 &  20.15849 & 19.84 & 19.20 & $  3.492 \pm  0.941$ & $  1.022 \pm  0.963$ &                      6 &              \\ 
SDSSJ02193313+2008302   & Segue 2 &  34.88806 &  20.14173 & 17.19 & 16.19 & $  1.694 \pm  0.188$ & $  0.128 \pm  0.121$ &                    6,8 &              \\ 
 DECJ02305767-7921156     & Hyi I &  37.74031 & -79.35435 & 19.03 & 18.44 & $  4.044 \pm  0.312$ & $ -1.755 \pm  0.276$ &                      9 &               \\
 DECJ02311155-7924240     & Hyi I &  37.79813 & -79.40667 & 19.81 & 19.29 & $  1.841 \pm  0.583$ & $  1.334 \pm  0.487$ &                      9 & PM non-member \\
OGL4J02294391-7916384     & Hyi I &  37.43306 & -79.27732 & 18.00 & 17.35 & $  3.963 \pm  0.236$ & $ -1.320 \pm  0.223$ &                      9 &               \\
OGL4J02323035-7927272     & Hyi I &  38.12667 & -79.45753 & 17.99 & 17.31 & $  3.545 \pm  0.298$ & $ -1.387 \pm  0.216$ &                      9 &               \\
\enddata
\tablenotetext{a}{References:  (1)  \citet{kirby15b};  (2)  \citet{martin16a};  (3)  \citet{kirby17};  (4)  \citet{venn17};  (5)  \citet{belokurov09};  (6)  \citet{kirby13};  (7)  \citet{boettcher13};  (8)  \citet{rk14};  (9)  \citet{koposov18};  (10)  \citet{koposov15b};  (11)  \citet{nagasaw18};  (12)  \citet{simon15};  (13)  \citet{ji16};  (14)  \citet{ji16c};  (15)  \citet{roederer16};  (16)  \citet{jf18};  (17)  \citet{li18};  (18)  \citet{torrealba18};  (19)  \citet{martin07};  (20)  \citet{sg07};  (21)  \citet{frebel10};  (22)  \citet{dallora12};  (23)  \citet{geha09};  (24)  \citet{norris10};  (25)  \citet{norris10c};  (26)  \citet{simon11};  (27)  \citet{fsk14};  (28)  \citet{kleyna05};  (29)  \citet{brown14};  (30)  \citet{siegel08};  (31)  \citet{willman11};  (32)  \citet{musella09};  (33)  \citet{koch09};  (34)  \citet{kr14};  (35)  \citet{ji16b};  (36)  \citet{francois16};  (37)  \citet{munoz06};  (38)  \citet{siegel06};  (39)  \citet{feltzing09};  (40)  \citet{norris10b};  (41)  \citet{koposov11};  (42)  \citet{lai11};  (43)  \citet{gilmore13};  (44)  \citet{ishigaki14};  (45)  \citet{frebel16};  (46)  \citet{martin16b};  (47)  \citet{walker16};  (48)  \citet{ji16d};  (49)  \citet{chiti18};  (50)  \citet{simon17};  (51)  \citet{hansen17}; }
\tablecomments{This table is available in its entirety in the electronic edition of the journal.  A portion is reproduced here to provide guidance on form and content.}
\label{mem_star_table}
\end{deluxetable*}

Even with only a handful of confirmed member stars in several of the
target galaxies (e.g., Hyi~1, Car~III, and Boo~II), the proper motions
are well-determined.  By combining \emph{Gaia} astrometry with
standard color selection, identification of just a few bright members
in a dwarf galaxy is sufficient to remove a very large fraction of the
Milky Way foreground contamination for spectroscopic follow-up.  I
illustrate this process using Hyi~I in Appendix~\ref{hyi1_select}.
Unsurprisingly, there are a few examples of stars that have been
spectroscopically classified as ultra-faint dwarf members in the
literature, but have proper motions very different from those of the
galaxies in which they supposedly reside.  In particular, several of
the spectroscopically-selected stars in Boo~I are astrometric
non-members (see Section~\ref{sec:boo1test}), one UMa~I star selected
as a member by multiple authors has a proper motion of 32~\masyr\ and
must be a foreground main-sequence star, and one of the four published
Hyi~I members also appears unlikely to be associated with the galaxy
based on its proper motion.  I remove these stars with deviant proper
motions before computing the weighted average.

\begin{figure}[th!]
\epsscale{1.2}
\plotone{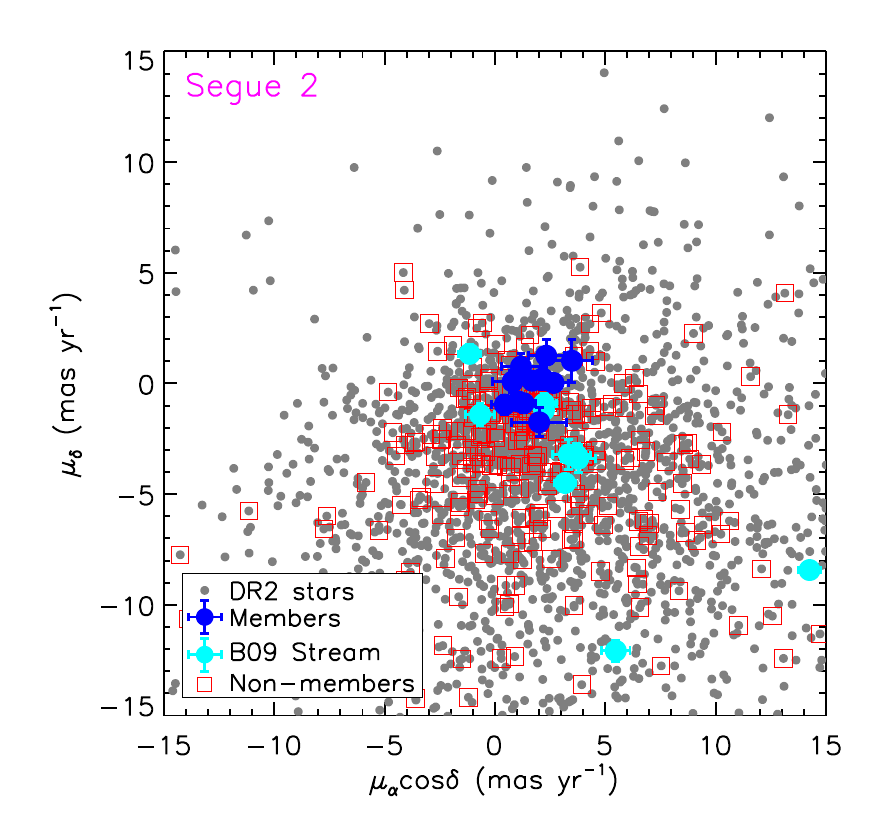}
\caption{Proper motion diagram for Segue~2.  Symbols are as in
  Figure~\ref{pmdiagrams}, with the exception of the cyan points,
  which represent members of the stream identified by
  \citet{belokurov09}.  }
\label{segue2_pm}
\end{figure}

The only system whose DR2 proper motions do not form an obvious
kinematically coherent group is Segue~2 (see Figure~\ref{segue2_pm}).
Two spectroscopic members of the \citet{belokurov09} stream have
proper motions (which are different from each other) above 10~\masyr,
so I reject them as being associated with Segue~2.  Even so, the
remainder of the distribution is broader than seen for the other
ultra-faint dwarfs.  The 8 candidate stream stars without wildly
deviant proper motions still exhibit a wide proper motion spread.
Four stars form a separate tight group with $\mua \sim 3$~\masyr,
$\mud \sim -3$~\masyr\, while the others are within or marginally
outside the distribution of Segue~2 stars.  These stars may be Segue 2
stars or field stars, but do not appear likely to be members of a
stream given their significantly different proper motions.  To be
conservative, I exclude all of the stream candidates from the Segue~2
proper motion sample.

The proper motions and corresponding space velocities in the Cartesian
Galactic frame\footnote{The sign conventions I use are that U
  velocities are positive toward the Galactic anticenter, V velocities
  are positive in the direction of Galactic rotation, and W velocities
  are positive toward the north pole of the Milky Way.  Note that this
  convention for U is opposite to that chosen by
  \citet{gaiadr2helmi}.} are presented in Table~\ref{pmdata_table}.
The space velocities of the ultra-faint dwarfs are surprisingly large,
exceeding 370~\kms\ with the exceptions of Segue~2, Willman~1, and
Tuc~III.  The orbits are also almost exclusively retrograde with
respect to Galactic rotation.  Only Draco~II and Tucana~III have any
significant motion in the direction of the rotation of the Milky Way
disk.  The vertical velocities are largely directed toward the north
pole of the Galaxy, with the exceptions being Segue~1, UMa~I, Boo~II,
and Dra~II, most of which are located at high Galactic latitude.  The
kinematics of the UFD population are qualitatively similar to those of
the Magellanic Clouds and Fornax, which also share positive W
velocities and highly negative V velocities \citep{gaiadr2helmi}.
This suggests a possible association of many of these dwarfs with the
Magellanic Clouds, as has been previously proposed
\citep{deason15,jethwa16,sales17}.

\begin{deluxetable*}{lcccccc}
\tablecaption{Gaia DR2 Proper Motions And Space Velocities}
\tablewidth{0pt}
\tablehead{
\colhead{Dwarf} &
\colhead{N$_{\rm stars}$} &
\colhead{$\mu_{\alpha}\cos{\delta}$} &
\colhead{$\mu_{\delta}$} & 
\colhead{U} &
\colhead{V} &
\colhead{W} \\
\colhead{} &
\colhead{} &
\colhead{(\masyr)} &
\colhead{(\masyr)} &
\colhead{(\kms)} &
\colhead{(\kms)} &
\colhead{(\kms)} 
}
\startdata
Triangulum~II & \phn2 & \phs$0.651 \pm 0.193$ &  \phs$0.592 \pm 0.164$ &  $-199.4$ & $-237.4$ &  \phs255.4 \\
Segue~2       & 11    & \phs$1.650 \pm 0.143$ &  $-0.065 \pm 0.094$ &    \phs157.6 & $-200.3$ &  \phs108.2 \\ 
Hydrus~I      & \phn3 & \phs$3.865 \pm 0.159$ &  $-1.450 \pm 0.135$ &    \phs180.4 & $-436.8$ &  \phs273.6 \\
Horologium~I  & \phn6 & \phs$0.901 \pm 0.070$ &  $-0.583 \pm 0.067$ & \phs\phn24.5 & $-426.1$ &  \phs162.6 \\
Reticulum~II  & 22    & \phs$2.393 \pm 0.040$ &  $-1.300 \pm 0.048$ & \phn\phn$-0.6$ & $-356.1$ &  \phs218.6 \\
Carina~II     & 19    & \phs$1.886 \pm 0.076$ &  \phs$0.079 \pm 0.070$ & $-119.1$ & $-544.8$ &  \phs147.1 \\
Carina~III    & \phn3 & \phs$3.035 \pm 0.120$ &  \phs$1.558 \pm 0.136$ & \phs\phn21.3 & $-402.5$ &  \phs347.3 \\
Ursa~Major~II & \phn8 & \phs$1.661 \pm 0.053$ &  $-1.870 \pm 0.065$ &    $-157.2$ & $-344.6$ &  \phs198.5 \\
Segue~1       & \phn9 & $-1.867 \pm 0.110$    &  $-3.282 \pm 0.102$ &   \phs132.7 & $-439.3$ &  $-48.7$ \\
Ursa~Major~I  & \phn8 & $-0.659 \pm 0.093$    &  $-0.635 \pm 0.131$ &   \phs197.3 & $-361.0$ & $-109.5$ \\ 
Willman~1     & \phn4 & \phs$0.382 \pm 0.119$ &  $-1.152 \pm 0.216$ &    $-109.8$ & $-211.4$ &  \phs102.2 \\
Coma~Berenices& 12    & \phs$0.546 \pm 0.092$ &  $-1.726 \pm 0.086$ &    $-262.3$ & $-251.0$ &   \phs\phn87.8 \\
Bo{\"o}tes~II     & \phn4 & $-2.517 \pm 0.325$    &  $-0.602 \pm 0.235$ & \phs$340.0$ & $-404.3$ &  \phn$-11.8$ \\
Bo{\"o}tes~I      & 68    & $-0.472 \pm 0.046$    &  $-1.086 \pm 0.034$  &   $-128.2$ & $-357.8$ &   \phs\phn56.6 \\
Draco~II      & \phn4 & \phs$1.170 \pm 0.297$ &  \phs$0.871 \pm 0.303$  & \phn\phn$-8.0$ & $-161.7$ & $-337.2$ \\
Tucana~II     & 10    & \phs$0.936 \pm 0.057$ &  $-1.232 \pm 0.072$ &   \phs259.0 & $-336.7$ &  \phs131.1 \\
Tucana~III    & 10    & $-0.014 \pm 0.038$    &  $-1.673 \pm 0.040$ &  \phn$-16.6$ & $-119.8$ &  \phs187.5 

\enddata
\label{pmdata_table}
\end{deluxetable*}

\begin{figure*}[th!]
\epsscale{1.2}
\plotone{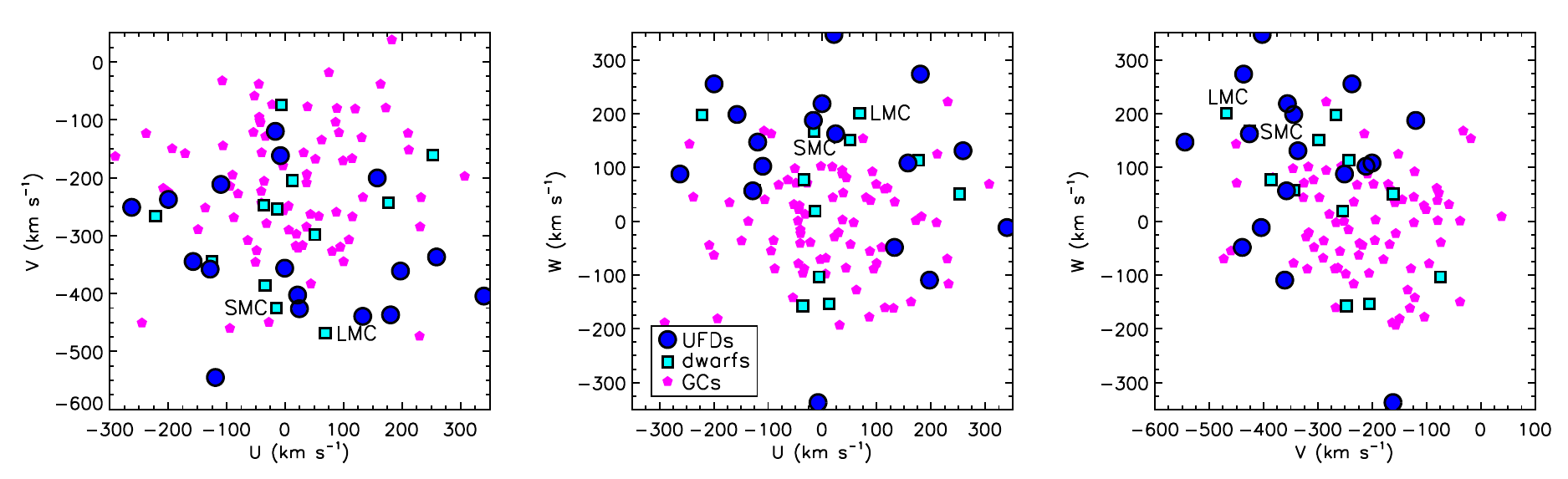}
\caption{UVW velocities for ultra-faint dwarfs (blue circles), more
  luminous dwarfs (cyan squares), and globular clusters (magenta
  pentagons).  The motions of the luminous dwarfs and globular
  clusters are taken from \citet{gaiadr2helmi}, adjusted for the
  different sign convention for U that I use.  The ultra-faint dwarfs
  are preferentially moving on retrograde orbits with positive
  vertical velocities, similar to the Magellanic Clouds.  The LMC
  $({\rm U}, {\rm V}, {\rm W}) = (68.6, -468.4, 201.0)$ and SMC $({\rm
    U}, {\rm V}, {\rm W}) = (-14.8, -425.0, 167.5)$ are labeled in
  each panel.}
\label{uvw}
\end{figure*}

\section{SATELLITE ORBITS AROUND THE MILKY WAY}
\label{sec:orbits}


Using the proper motions determined with \emph{Gaia} in Section~\ref{sec:pms}
and positions and radial velocities from the literature, I calculate
the orbit of each satellite around the Milky Way with the
\texttt{galpy} software package \citep{galpy}.  For simplicity and to
maximize the reproducibility of the results, I use the recommended
MWPotential2014 gravitational potential for the Milky Way with its
default parameters ($R_{0} = 8.0$~kpc, $V_{0} = 220$~\kms); I note
that there is not currently compelling observational evidence to favor
the adoption of a different potential.  I assume the solar motion
determined by \citet{schoenrich10} throughout this paper.

Because the observational uncertainties on the proper motions and
distances are large in some cases (see Tables~\ref{litdata_table} and
\ref{pmdata_table}), I calculate orbits by drawing 1000 random proper
motion, distance and radial velocity values from Gaussian
distributions.  The Gaussian distributions are defined such that the
mean is the measured value of that parameter, and the standard
deviation is equal to the $1\sigma$ observational uncertainty.  I
recompute the orbit for each of the 1000 parameter sets and take the
16th, 50th, and 84th percentiles of the resulting distributions to
describe the median and $1\sigma$ confidence interval of the orbit.
Uncertainties on the central positions of the ultra-faint dwarfs have
only been published for about half of the objects included in my
sample.  The typical positional uncertainties for the systems for
which they have been reported are $\sim0.003\degr$ ($\sim11\arcsec$)
per coordinate.  These uncertainties impart negligible changes to the
orbits, so I ignore them.  I report the derived orbital pericenters,
apocenters, eccentricities, and periods in Table~\ref{orbit_table},
and projections of the orbits are shown in Figures~\ref{orbits_inner}
and \ref{orbits_outer}.

\begin{deluxetable}{lllcc}
\tablecaption{Orbits}
\tablewidth{0pt}
\tablehead{
\colhead{Dwarf} &
\colhead{Pericenter} &
\colhead{Apocenter} & 
\colhead{Eccentricity} &
\colhead{Period} \\
\colhead{} &
\colhead{(kpc)} &
\colhead{(kpc)} &
\colhead{} &
\colhead{(Gyr)} 
}
\startdata
 Triangulum~II &  $20^{+2}_{-3}$ & $541^{+493}_{-211}$ & $0.93^{+0.03}_{-0.03}$ & $>$10 \\
       Segue~2 &  $32^{+4}_{-5}$ & $54^{+14}_{-7}$ & $0.32^{+0.62}_{-0.04}$ & $1.1$ \\ 
      Hydrus~I &  $26^{+1}_{-1}$ & $1029^{+642}_{-610}$ & $0.95^{+0.02}_{-0.06}$ & $>$10 \\  
  Horologium~I &  $87^{+12}_{-15}$ & $741^{+1331}_{-619}$ & $0.79^{+0.12}_{-0.50}$ & $...$ \\  
  Reticulum~II &  $29^{+4}_{-6}$ & $91^{+91}_{-39}$ & $0.51^{+0.18}_{-0.13}$ & $1.6$ \\   
     Carina~II &  $28^{+1}_{-1}$ & $1521^{+232}_{-231}$ & $0.96^{+0.01}_{-0.01}$ & $>$10 \\ 
    Carina~III &  $29^{+1}_{-1}$ & $1478^{+436}_{-488}$ & $0.96^{+0.01}_{-0.02}$ & $>$10 \\ 
 Ursa~Major~II &  $39^{+2}_{-3}$ & $201^{+179}_{-80}$ & $0.67^{+0.13}_{-0.14}$ & $3.5$ \\ 
       Segue~1 &  $20^{+4}_{-5}$ & $61^{+34}_{-18}$ & $0.52^{+0.08}_{-0.03}$ & $1.1$ \\ 
  Ursa~Major~I &  $102^{+6}_{-7}$ & $896^{+937}_{-669}$ & $0.79^{+0.1}_{-0.39}$ & $...$ \\ 
     Willman~1 &  $44^{+15}_{-19}$ & $53^{+57}_{-13}$ & $0.19^{+0.19}_{-0.11}$ & $1.2$ \\ 
Coma~Berenices &  $43^{+2}_{-2}$ & $242^{+192}_{-87}$ & $0.70^{+0.12}_{-0.12}$ & $3.7$ \\ 
     Bo{\"o}tes~II &  $39^{+2}_{-2}$ & $1746^{+1439}_{-1120}$ & $0.96^{+0.02}_{-0.17}$ & $>$10 \\ 
      Bo{\"o}tes~I &  $45^{+5}_{-5}$ & $116^{+21}_{-15}$ & $0.44^{+0.03}_{-0.01}$ & $2.4$ \\ 
      Draco~II &  $18^{+3}_{-3}$ & $158^{+255}_{-71}$ & $0.79^{+0.11}_{-0.10}$ & $2.0$ \\ 
     Tucana~II &  $39^{+12}_{-12}$ & $626^{+1654}_{-449}$ & $0.78^{+0.16}_{-0.09}$ & $...$ \\ 
    Tucana~III &  $3^{+1}_{-1}$ & $49^{+6}_{-5}$ & $0.90^{+0.01}_{-0.01}$ & $0.7$ \\ 
\enddata
\label{orbit_table}
\end{deluxetable}

In addition to Boo~I (Section~\ref{sec:boo1test}), the other
ultra-faint dwarf with a literature proper motion measurement is
Segue~1.  \cite{fritz17} determined a proper motion of $\mua = -0.37
\pm 0.57$~\masyr, $\mud = -3.39 \pm 0.58$~\masyr\ using ground-based
data over a time baseline of 10~years.  The DR2 measurement has
uncertainties smaller by a factor of $\sim5$ and is in excellent
agreement in the Declination direction, but the motions in Right
Ascension differ by $2.6\sigma$.  Nevertheless, the derived orbits are
quite similar, with overlapping pericenter and apocenter ranges.
Although the proper motion of Tuc~III has not previously been
measured, \citet{erkal18} predicted its proper motion based on the
stellar kinematics of its tidal tails provided by \citet{li18b}.  The
properties of the Tuc~III orbit such as its eccentricity and
pericenter, as well as the $\mud$ value, are close to the
\citeauthor{erkal18} prediction, but the proper motion in $\mua$ is
much closer to zero than expected.  Exploration of the Tuc~III orbit
in different gravitational potentials for the Milky Way and the LMC
may be required to understand this discrepancy.

The pericentric distances of the ultra-faint dwarfs are tightly
constrained by the \emph{Gaia} proper motions, with typical
uncertainties of 5~kpc.  The best-fit pericenter is beyond 20~kpc for
every galaxy except Tuc~III, which approaches within 3~kpc of the
Galactic Center.  Not surprisingly given this extreme orbit, Tuc~III
has been tidally disrupted, with obvious tidal tails in DES imaging
\citep{drlica15,shipp18} and a kinematic gradient detected along the
tails \citep{li18b,erkal18}.  None of the other dwarfs have orbits
that penetrate significantly into the disk or bulge.  Strikingly, a a
large majority of these dwarfs are currently located very close to the
pericenters of their orbits around the Milky Way.  Whether this result
can be attributed to observational selection effects or other factors
is not yet clear.

Under the assumption that the mass distribution of the Milky Way
follows that of the MWPotential2014 model, nearly half of the target
galaxies lack well-determined orbital apocenters, and they appear
likely to be on their first orbit around the Milky Way.  However, the
MWPotential2014 mass model corresponds to a relatively low-mass Milky
Way of M$_{\rm vir} = 0.8 \times 10^{12}$~M$_{\odot}$ \citep{galpy}.
The large fraction of objects apparently on their first infall to the
Milky Way when their orbits are integrated in this potential suggests
that the mass of the Milky Way might be underestimated in this model.
To test this idea, I recomputed orbits for the eight dwarfs with very
large apocenters in the extreme case of a Milky Way circular velocity
of 250~\kms, compared to the assumed value of 220~\kms.  With this
more massive potential, seven of the eight first-infall objects have
orbital apocenters of less than 300~kpc and typical periods of
$2-4$~Gyr.  In some fraction of the Monte Carlo iterations these
dwarfs are still on their first infall, but such orbits become a
minority of those simulated.  The only exception is Boo~II, which
still has an apocenter of $482^{+1592}_{-347}$~kpc in this potential
and is on its first infall.


\section{DISCUSSION}
\label{sec:discussion}

\subsection{Tidal Stripping of the Ultra-Faint Dwarfs}

Since shortly after their initial discovery, it has frequently been
speculated that many ultra-faint dwarfs are heavily tidally stripped
systems
\citep[e.g.,][]{no09,mj10,munoz10,sand12,kirby13,roderick15,roderick16,collins17,garling18}.
On the other hand, \citet{penarrubia08} showed using N-body
simulations that the properties of the ultra-faint dwarfs did not seem
to be consistent with being the tidally-stripped remnants of the
classical dSphs.  The existence of a mass-metallicity relationship
extending down to the lowest luminosity dwarfs is also inconsistent
with the idea that many of the ultra-faint dwarfs have been
significantly stripped \citep{kirby13b}.  However, only orbital
measurements can show conclusively whether these systems have been
subject to strong enough tidal forces to remove many of their stars.
For Tuc~III, the answer is unequivocally yes, as shown by its orbit
derived here and the studies of \citet{li18b} and \citet{erkal18}.
The other dwarfs in my sample, though, remain well away from the
regions where the Milky Way's tidal field is strongest.  After
Tuc~III, the next closest orbital pericenters are those of Segue~1 and
Tri~II, each at 20~kpc.  Using the mass determined from the internal
kinematics of its stars, \citet{simon11} showed that the tidal radius
of Segue~1 remains well beyond its stellar distribution unless the
galaxy has an orbital pericenter of $\sim4$~kpc or less.  My results
here show that Segue~1 cannot have experienced significant stellar
stripping (also see \citealt{fritz17} for independent confirmation).
The same is likely true for most of the rest of the ultra-faint dwarf
population.  While the dark matter halos of these galaxies, which
extend to much larger radii than the stars, may suffer substantial
tidal effects, the stars at the center of the halo are protected from
the influence of Galactic tides.

\begin{figure*}[t!]
\includegraphics[width=6cm]{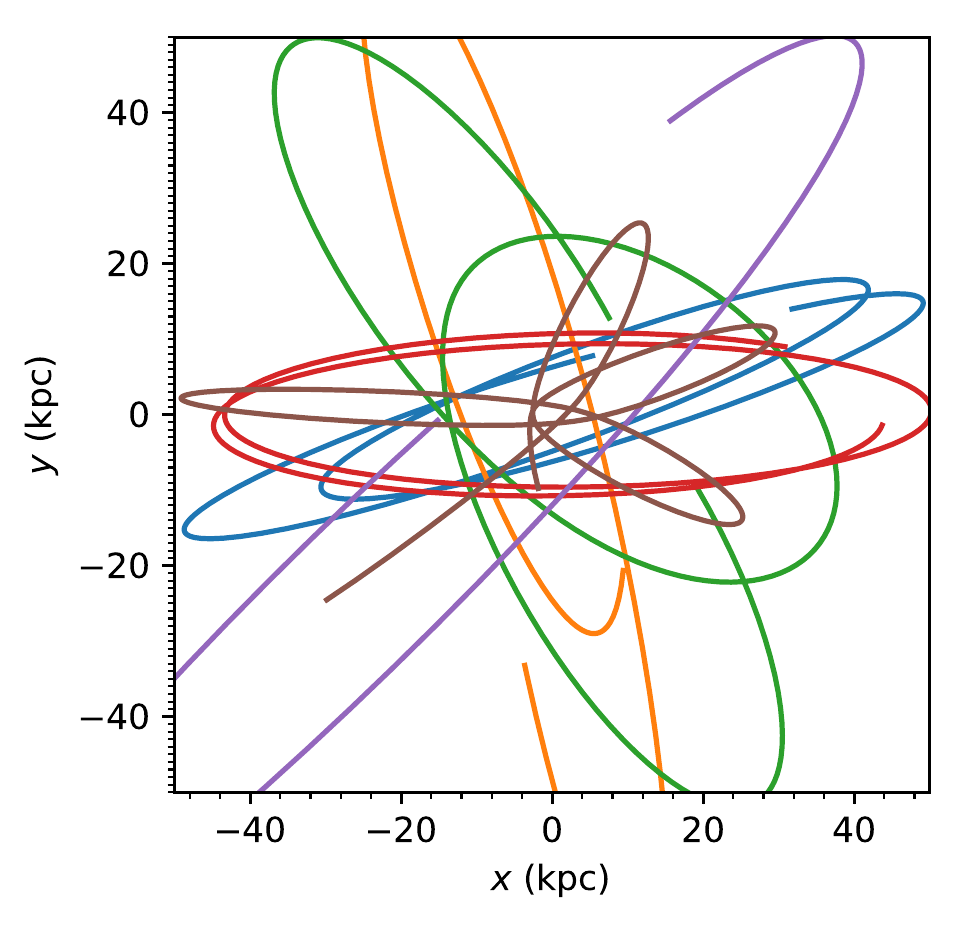}\includegraphics[width=6cm]{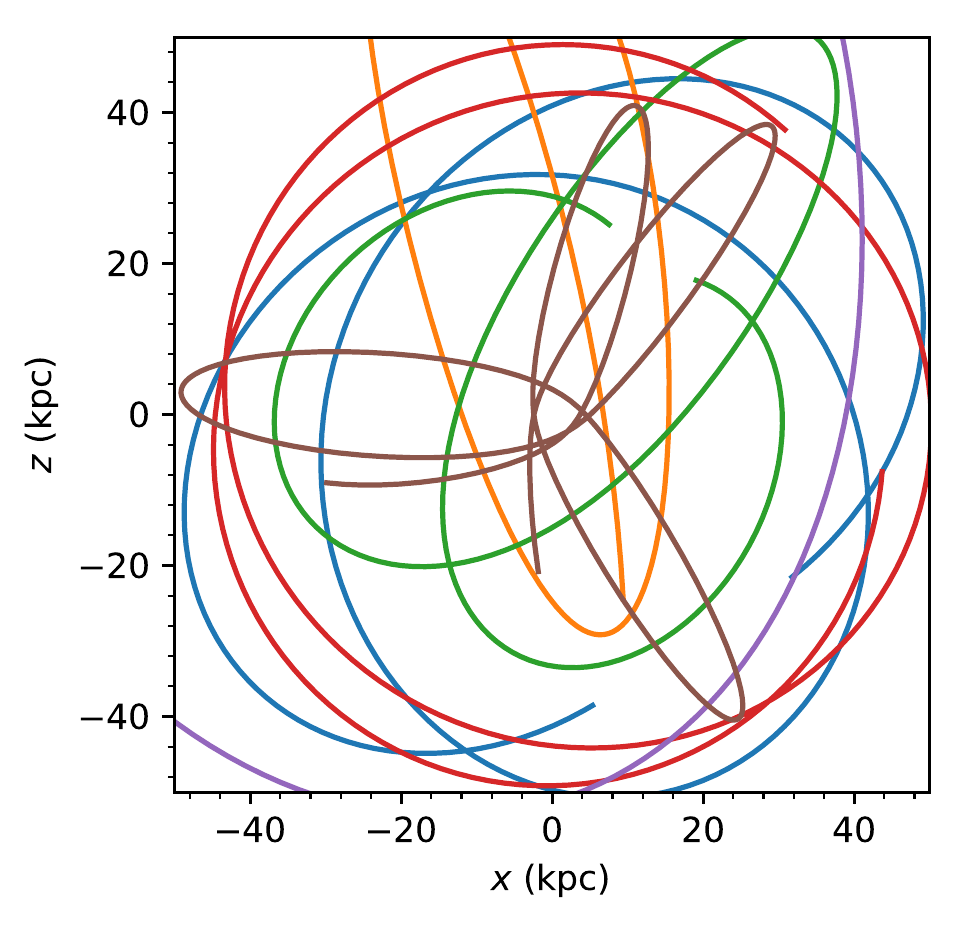}\includegraphics[width=6cm]{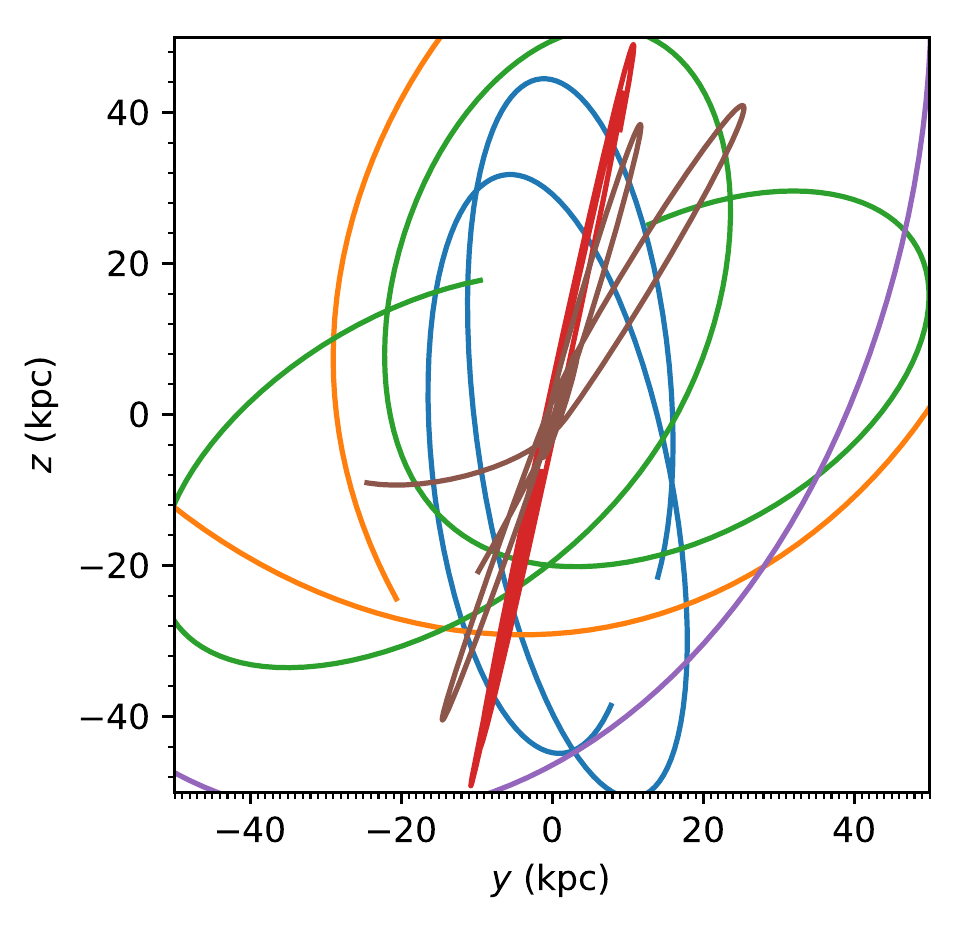}
\caption{(left) Projections of the orbits on the X-Y plane for the
  ultra-faint dwarfs that are confined to Galactocentric radii of less
  than 100~kpc. (middle) Projections of the orbits on the X-Z plane
  for the same dwarfs. (right) Projections of the orbits on the Y-Z
  plane.  Segue~2 is plotted as the blue curve, Ret~II is orange,
  Segue~1 is green, Willman~1 is red, Boo~I is purple, and Tuc~III is
  brown.  The orbits are integrated for 4~Gyr.}
\label{orbits_inner}
\end{figure*}

\begin{figure*}[th!]
\includegraphics[width=6cm]{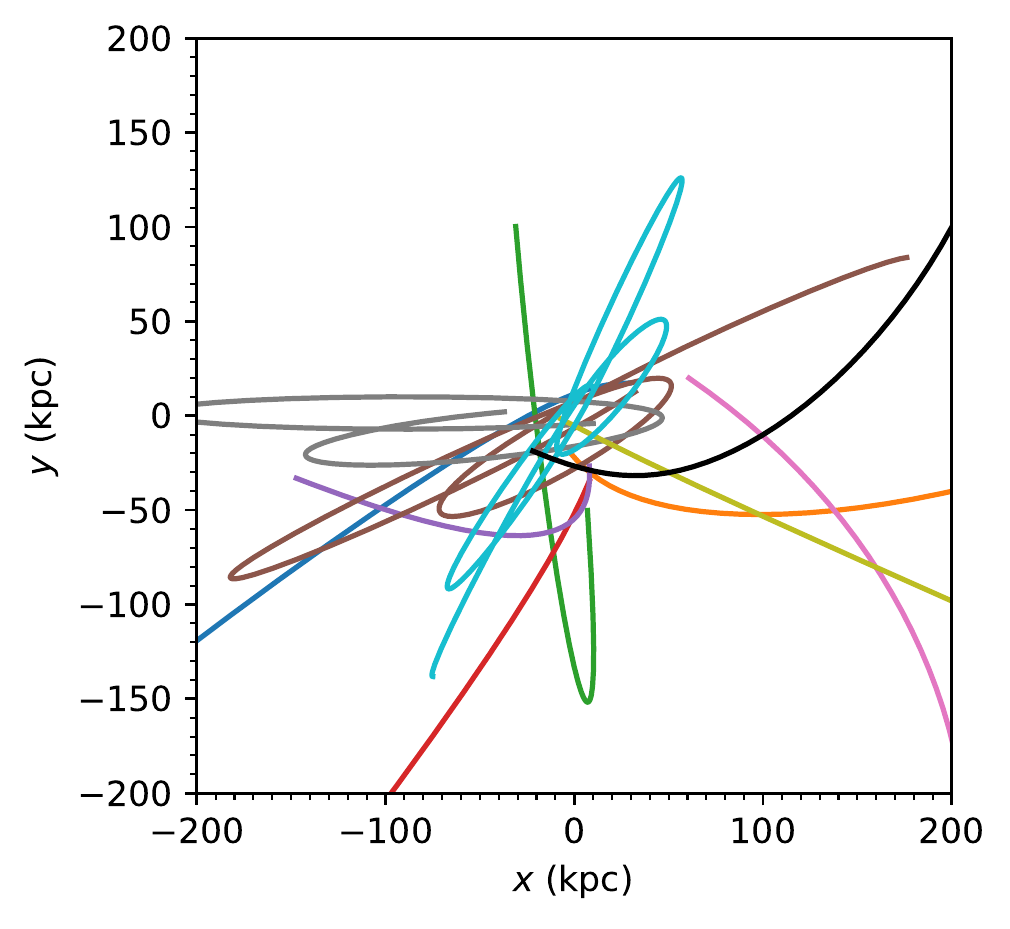}\includegraphics[width=6cm]{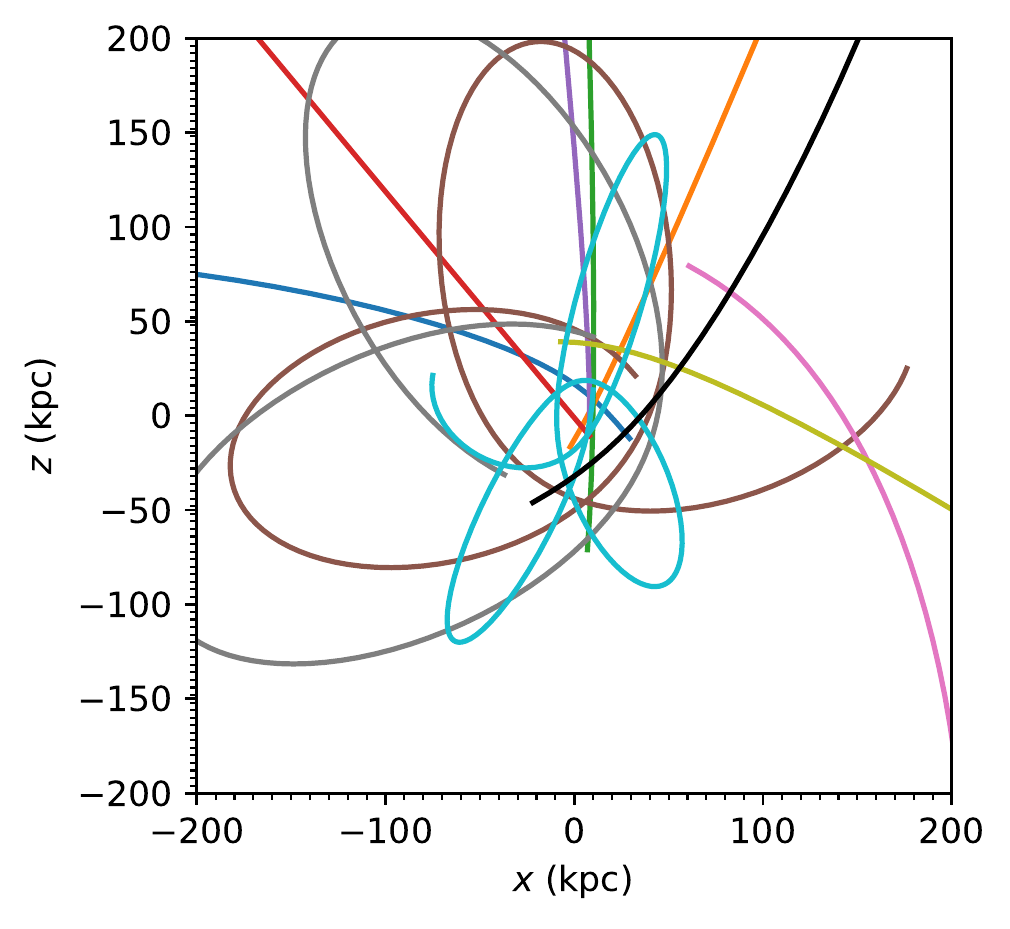}\includegraphics[width=6cm]{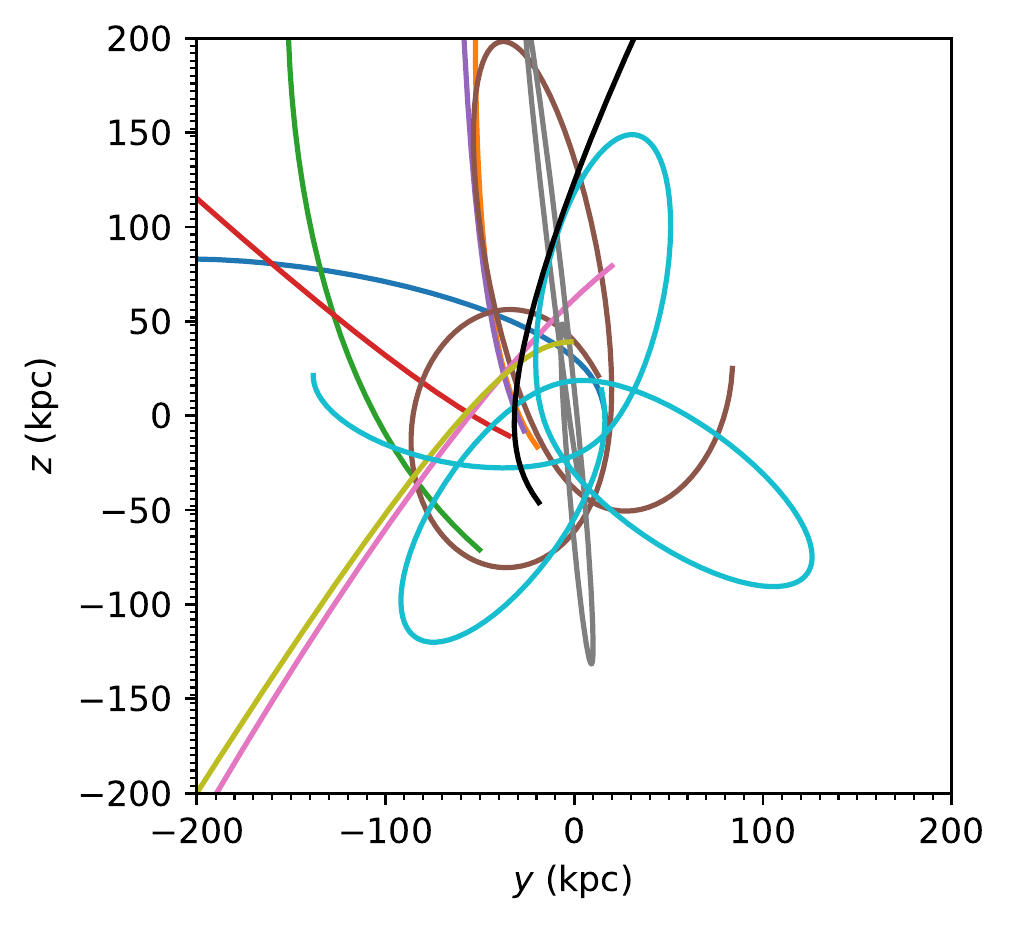}
\caption{(left) Projections of the orbits on the X-Y plane for the
  ultra-faint dwarfs with orbits that extend beyond 200~kpc. (middle)
  Projections of the orbits on the X-Z plane for the same
  dwarfs. (right) Projections of the orbits on the Y-Z plane.  Tri~II
  is plotted as the blue curve, Hyi~I is orange, Hor~I is green,
  Car~II is red, Car~III is purple, UMa~II is brown, UMa~I is pink,
  Com~Ber is gray, Boo~II is chartreuse, Dra~II is cyan, and Tuc~II is
  black.  The orbits are integrated for 10~Gyr.}
\label{orbits_outer}
\end{figure*}

\subsection{Comparison to Predicted Proper Motions}

Several recent studies have provided predictions for the proper
motions of the ultra-faint dwarfs, based either on the possibility
that they are distributed in a thin, rotating vast polar structure
\citep[VPoS;][]{pawlowski13,pawlowski15} or that they were accreted as
members of a Magellanic group of dwarf galaxies
\citep{jethwa16,sales17}.  The results of the \citet{pawlowski13} and
\citet{pawlowski15} predictions are mixed: several satellites are
consistent with being co-rotating with the polar structure (Hor~I,
Ret~II, Willman~1, and Tuc~II), Segue~1 is counter-rotating, and four
other dwarfs (UMa~II, Com~Ber, Boo~II, and Boo~I) are outside the
expected proper motion ranges.  Moreover, the orbit integrations shown
in Figures~\ref{orbits_inner} and \ref{orbits_outer} indicate that the
population as a whole is not confined to a thin plane, consistent with
the results of \citet{gaiadr2helmi} for the more luminous dwarf
galaxies.


\citet{jethwa16} provide predictions for the proper motions of four of
the targets of this study under the assumption that they originated as
satellites of the Magellanic Clouds.  As noted earlier, there are
qualitative resemblances between many of the orbits I derive and those
of the Magellanic Clouds.  Hor~I and Ret~II have proper motions in the
right ascension direction in agreement with the \citeauthor{jethwa16}
predictions, but the motions in declination are inconsistent with the
predicted values at the $2-3\sigma$ level.  For Tuc~II, the opposite
is the case: the proper motion in declination is marginally consistent
with the prediction, but the proper motion in R.A. is not.  The
predicted ranges for Tuc~III are so wide that the satellite was
virtually guaranteed to fall within them.  These four objects, as well
as Hyi~I, Car~II, and Car~III, all of which were discovered too
recently for predictions to be available, can be regarded as candidate
Magellanic satellites.  Viewed in three dimensions and integrated
backward in time, the orbits of Hor~I, Hyi~I, Car~III, and Tuc~II
track that of the LMC quite closely, suggesting that these objects are
probably associated with the Magellanic system.  \citet{deason15} and
\citet{sales17} both consider Hor~I and Tuc~II to be the most likely
dwarfs of Magellanic origin, which is consistent with the results
reported here.

\subsection{Comparison to Numerical Simulations}

N-body simulations of dark matter halo formation in the $\Lambda$CDM
paradigm show that subhalos are usually on quite radial orbits, with
typical pericenter to apocenter ratios of $\sim0.2$
\citep[e.g.,][]{vandenbosch99,diemand07,wetzel11,barber14}.  Contrary
to this expectation, \citet{gaiadr2helmi} find that the orbits of the
classical dSphs are relatively round, with the exceptions of the most
distant objects, Leo~I and Leo~II.  On the other hand, the results
presented here are very consistent with the predicted orbital
distribution, with eccentricities exceeding 0.9 for more than a third
of the sample.  The large eccentricities I measure may be connected to
the preference for first-infall orbits discussed in
Section~\ref{sec:orbits}; \citet{gill04} showed that satellite orbits
tend to get rounder as their time within the host halo increases.

It is worth noting that the N-body results may not represent an
accurate prediction for what should be observed.  In simulations
including hydrodynamics, many of the subhalos that approach within
30~kpc of the center of their host galaxy are destroyed by tidal
interactions \citep{gk17,sawala17}.  In that respect, it may be
unexpected that so many of the dwarfs have orbital pericenters of less
than 30~kpc.  However, because their orbits are so eccentric they
spend very little time in the central regions of the Milky Way,
enabling them to survive until the present \citep{sawala17}.  More
detailed work on the predicted distribution of subhalo orbits when
baryonic physics is included in simulations would be fruitful now that
there is a significant set of observations with which to compare.

\subsection{Infall Times}

\citet{rocha12} showed that there is a strong correlation between the
binding energy of a subhalo and the time it was accreted by the host
halo.  Using Eq.~1 from their paper, I calculate the binding energies
for the 17 ultra-faint dwarfs studied here.  For the majority of the
sample the binding energy defined in this way is negative and the
\citet{rocha12} analysis does not apply.  This result is obviously
related to the apparent first-infall status of many of these galaxies.
Among the galaxies for which the binding energy is well-behaved, the
\citet{rocha12} relation suggests infall times of $\sim7$~Gyr for
Tri~II, Segue~2, and Willman~1, $\sim8$~Gyr for Draco~II and Tuc~III,
and $>12$~Gyr for Ret~II.  Ret~II has a higher binding energy than any
of the Via~Lactea~2 subhalos included in the \citeauthor{rocha12}
sample.

\subsection{Implications for the Mass of the Milky Way}

As discussed in previous sections and shown in
Figure~\ref{orbits_outer}, a number of the target galaxies of this
study appear to be very loosely bound and may be on their first
orbital passage around the Milky Way, which would be an unexpected
result.  I showed in Section~\ref{sec:orbits} that this problem can be
alleviated if the mass of the Milky Way is larger than assumed in the
default \texttt{galpy} MWPotential2014 model.  Relatively large Milky Way
masses have been advocated by some authors
\citep[e.g.,][]{boylan-kolchin13}, but if the mass of the Galaxy is
large that will tend to exacerbate issues such as the missing
satellite problem \citep{klypin99,moore99} and the too big to fail
problem \citep{bk11,bk12}.

The alternative explanation is that many of the ultra-faint satellites
are on their first orbit because they formed in a Magellanic group
that is now infalling into the Milky Way for the first time
\citep{besla07,kallivayalil13}.  However, if many of the currently
known satellites were originally located around the LMC, that also has
adverse consequences for the missing satellite problem, because the
number of dwarfs formed around the Milky Way would be even smaller.
Improved estimates of the Milky Way's mass based on dynamical modeling
of the kinematics of the satellite population provided here and by
\citet{gaiadr2helmi} will be valuable for attempting to resolve this
dilemma.

\section{SUMMARY AND CONCLUSIONS}
\label{sec:conclusions}

I have presented \emph{Gaia} DR2-based proper motions, space velocities, and
Galactic orbits for 17 ultra-faint satellites of the Milky Way.  I
cross-match lists of member stars that are confirmed via spectroscopy
or variable star pulsations with the \emph{Gaia} DR2 catalog to determine the
motion of each system across the sky.  Thanks to the exquisite
astrometric performance of \emph{Gaia}, even with as few as $\sim3$ member
stars brighter than $g=20$ per satellite, the proper motions are
determined with a typical precision of $\sim0.1$~\masyr\ per
coordinate.  At a median distance of 35~kpc, the uncertainties on the
tangential velocities are therefore $\sim25$~\kms.  I find good
agreement with the proper motion measurements reported by
\citet{gaiadr2helmi} for Bo{\"otes}~I and Leo~I, as well as with
\emph{HST} measurements of Leo~I \citep{sohn13} and $\Omega$~Centauri
\citep{libralato18}.

The tangential velocities of the sample of ultra-faint dwarfs analyzed
here are generally much larger than the line-of-sight velocities
measured in the literature, with a median value exceeding 400~\kms.
Most of the dwarfs are orbiting in the opposite direction from the
rotation of the Milky Way and are moving toward the northern
hemisphere of the Galaxy, broadly similar to the space motions of the
Magellanic Clouds.  I use \texttt{galpy} \citep{galpy} to integrate
the orbits of the ultra-faint dwarfs in the potential of the Milky
Way.  The resulting orbital parameters are surprising in a number of
related ways: 

\begin{itemize}

\item The dwarfs are moving at very high velocities.  The tangential
  velocities are significantly higher than the radial velocities for
  14 out of the 17 dwarfs.

\item A large fraction of the satellites are currently located near
  the pericenters of their orbits.  More than half of the sample
  (Hyi~I, Hor~I, Ret~II, Car~III, UMa~II, UMa~I, Com~Ber, Boo~II, and
  Dra~II) is within 4~kpc of the derived pericenter, and most of the
  satellites are no more than 100~Myr away from passing or reaching
  pericenter.

\item The orbital eccentricities are very high, comparable to those
  found in pure dark matter N-body simulations, but significantly
  larger than those of the classical dwarf spheroidals
  \citep{gaiadr2helmi}.  Only Willman~1 is robustly on a circular
  orbit; the orbit of Segue~2 is consistent with being relatively
  round but with large uncertainties.  Segue~1 and Boo~I have orbital
  eccentricities of $\sim0.5$, and the remainder of the sample has
  eccentricities ranging from 0.67 to 0.96.

\item If the recommended \texttt{galpy} gravitational potential for
  the Milky Way is used, many of the ultra-faint dwarfs have orbital
  apocenters well beyond the virial radius of the Milky Way and have
  yet to complete a single orbit around the Galaxy.  A larger mass for
  the Milky Way can bring the derived apocenters down to more
  reasonable values (with the exception of Boo~II), at the cost of
  worsening other problems in near-field cosmology and increasing
  tension with a variety of evidence for a relatively low-mass Milky
  Way.  Alternatively, perhaps more of these satellites than expected
  have been stripped from the Magellanic system.

\end{itemize}

Of the galaxies considered in this paper, only the tidally disrupting
Tuc~III system has an orbital pericenter of less than 20~kpc.  Because
of their eccentric orbits and large pericentric distances, the stellar
components of the other dwarfs likely have not been significantly
affected by Galactic tides.

Predictions for the proper motions of a number of the dwarfs studied
here have been made in the literature, assuming either that they are
orbiting within a vast polar structure \citep{pawlowski13,pawlowski15}
or that they were accreted by the Milky Way from the Magellanic Clouds
\citep{jethwa16,sales17}.  A portion of the sample has velocities that
are consistent with the VPoS predictions, but other objects do not,
and the orbits of the full sample are not primarily confined to a thin
plane.  Several of the southern dwarfs, such as Hyi~I, Hor~I, Car~III,
and Tuc~II, have orbits suggesting that they are (or were) Magellanic
satellites.

The \emph{Gaia} DR2 data represent such an enormous step forward in size,
depth, and astrometric accuracy that the analysis I present here only
scratches the surface of what is now possible.  \emph{Gaia} measurements can
likely be obtained for additional dwarf galaxies, bringing the
dynamics of the Milky Way's entire surroundings into sharp focus for
the first time.  The measured proper motions for many of the dwarfs
studied here can also be improved by obtaining larger spectroscopic
(or photometric) member samples, and the efficiency of spectroscopic
target selection will be greatly enhanced by the availability of \emph{Gaia}
astrometry once the approximate proper motion of any nearby dwarf is
known.  I look forward to these and many other foreseen and unforeseen
results in the near future.  The \emph{Gaia} revolution has begun!

\acknowledgements{I thank Marla Geha, Juna Kollmeier, Sal Fu, Ting Li,
  Jo Bovy, Alex Ji, Andrew Wetzel, and Denis Erkal for helpful
  conversations.  I also thank the entire \emph{Gaia} collaboration for their
  hard work in producing this astonishing data set.  This publication
  is based upon work supported by the National Science Foundation
  under grant AST-1714873.  Assemblage of the spectroscopic samples
  for some of the dwarf galaxies included here was supported by a NASA
  Keck PI Data Award, administered by the NASA Exoplanet Science
  Institute under RSA number 1474359.  This research has made use of
  NASA's Astrophysics Data System Bibliographic Services.

  This work has made use of data from the European Space Agency (ESA)
  mission {\it Gaia} (\url{https://www.cosmos.esa.int/gaia}),
  processed by the {\it Gaia} Data Processing and Analysis Consortium
  (DPAC,
  \url{https://www.cosmos.esa.int/web/gaia/dpac/consortium}). Funding
  for the DPAC has been provided by national institutions, in
  particular the institutions participating in the {\it Gaia}
  Multilateral Agreement.

  This publication made use of data from the Sloan Digital Sky Survey.
  Funding for SDSS-III has been provided by the Alfred P. Sloan
  Foundation, the Participating Institutions, the National Science
  Foundation, and the U.S. Department of Energy Office of Science. The
  SDSS-III web site is http://www.sdss3.org/.

  SDSS-III is managed by the Astrophysical Research Consortium for the
  Participating Institutions of the SDSS-III Collaboration including
  the University of Arizona, the Brazilian Participation Group,
  Brookhaven National Laboratory, Carnegie Mellon University,
  University of Florida, the French Participation Group, the German
  Participation Group, Harvard University, the Instituto de
  Astrof{\'i}sica de Canarias, the Michigan State/Notre Dame/JINA
  Participation Group, Johns Hopkins University, Lawrence Berkeley
  National Laboratory, Max Planck Institute for Astrophysics, Max
  Planck Institute for Extraterrestrial Physics, New Mexico State
  University, New York University, Ohio State University, Pennsylvania
  State University, University of Portsmouth, Princeton University,
  the Spanish Participation Group, University of Tokyo, University of
  Utah, Vanderbilt University, University of Virginia, University of
  Washington, and Yale University.

  This project used public archival data from the Dark Energy Survey
  (DES). Funding for the DES Projects has been provided by the
  U.S. Department of Energy, the U.S. National Science Foundation, the
  Ministry of Science and Education of Spain, the Science and
  Technology FacilitiesCouncil of the United Kingdom, the Higher
  Education Funding Council for England, the National Center for
  Supercomputing Applications at the University of Illinois at
  Urbana-Champaign, the Kavli Institute of Cosmological Physics at the
  University of Chicago, the Center for Cosmology and Astro-Particle
  Physics at the Ohio State University, the Mitchell Institute for
  Fundamental Physics and Astronomy at Texas A\&M University,
  Financiadora de Estudos e Projetos, Funda{\c c}{\~a}o Carlos Chagas
  Filho de Amparo {\`a} Pesquisa do Estado do Rio de Janeiro, Conselho
  Nacional de Desenvolvimento Cient{\'i}fico e Tecnol{\'o}gico and the
  Minist{\'e}rio da Ci{\^e}ncia, Tecnologia e Inova{\c c}{\~a}o, the
  Deutsche Forschungsgemeinschaft, and the Collaborating Institutions
  in the Dark Energy Survey.

  The Collaborating Institutions are Argonne National Laboratory, the
  University of California at Santa Cruz, the University of Cambridge,
  Centro de Investigaciones Energ{\'e}ticas, Medioambientales y
  Tecnol{\'o}gicas-Madrid, the University of Chicago, University
  College London, the DES-Brazil Consortium, the University of
  Edinburgh, the Eidgen{\"o}ssische Technische Hochschule (ETH)
  Z{\"u}rich, Fermi National Accelerator Laboratory, the University of
  Illinois at Urbana-Champaign, the Institut de Ci{\`e}ncies de
  l'Espai (IEEC/CSIC), the Institut de F{\'i}sica d'Altes Energies,
  Lawrence Berkeley National Laboratory, the Ludwig-Maximilians
  Universit{\"a}t M{\"u}nchen and the associated Excellence Cluster
  Universe, the University of Michigan, the National Optical Astronomy
  Observatory, the University of Nottingham, The Ohio State
  University, the OzDES Membership Consortium, the University of
  Pennsylvania, the University of Portsmouth, SLAC National
  Accelerator Laboratory, Stanford University, the University of
  Sussex, and Texas A\&M University.

  Based in part on observations at Cerro Tololo Inter-American
  Observatory, National Optical Astronomy Observatory, which is
  operated by the Association of Universities for Research in
  Astronomy (AURA) under a cooperative agreement with the National
  Science Foundation.

  This publication made use of data from Pan-STARRS DR1.  The
  Pan-STARRS1 Surveys (PS1) and the PS1 public science archive have
  been made possible through contributions by the Institute for
  Astronomy, the University of Hawaii, the Pan-STARRS Project Office,
  the Max-Planck Society and its participating institutes, the Max
  Planck Institute for Astronomy, Heidelberg and the Max Planck
  Institute for Extraterrestrial Physics, Garching, The Johns Hopkins
  University, Durham University, the University of Edinburgh, the
  Queen's University Belfast, the Harvard-Smithsonian Center for
  Astrophysics, the Las Cumbres Observatory Global Telescope Network
  Incorporated, the National Central University of Taiwan, the Space
  Telescope Science Institute, the National Aeronautics and Space
  Administration under Grant No. NNX08AR22G issued through the
  Planetary Science Division of the NASA Science Mission Directorate,
  the National Science Foundation Grant No. AST-1238877, the
  University of Maryland, E{\"o}tv{\"o}s Lor{\'a}nd University (ELTE),
  the Los Alamos National Laboratory, and the Gordon and Betty Moore
  Foundation.

  This research made use of NumPy \citep{van2011numpy}, matplotlib, a
  Python library for publication quality graphics \citep{Hunter:2007},
  and Astropy, a community-developed core Python package for Astronomy
  \citep{astropy}.

}

\bibliographystyle{apj}

\begin{appendix}

\section{Triangulum II}

The first spectroscopy of Triangulum~II (Tri~II) was obtained by
\citet{kirby15b} with Keck/DEIMOS, which was quickly followed up with
additional DEIMOS spectroscopy by \citet{martin16a}.  Given the
discrepant results between the two studies, \citet{kirby17} reobserved
the entire sample of member stars identified by \citet{kirby15b} and
\citet{martin16a} with DEIMOS.  The chemical abundances of the
brightest two stars on the red giant branch were studied at high
spectral resolution with Gemini-N/GRACES by \citet{venn17}, and a
similar analysis of the brighter star was carried out by
\citet{kirby17} with Keck/HIRES.  Because of its very low luminosity,
Tri~II contains just two spectroscopically confirmed member stars
brighter than $g=20$, but the three DEIMOS studies produced a total of
32 confirmed non-members brighter than this limit.

\section{Segue 2}

\citet{belokurov09} presented MMT/Hectochelle spectroscopy of Segue~2
in the discovery paper for the system, which \citet{kirby13}
supplemented with a large Keck/DEIMOS data set.  Imaging of Segue~2 by
\citet{boettcher13} revealed 1 RR~Lyrae variable, which was also
identified spectroscopically by \citet{kirby13}.  \citet{rk14}
analyzed the chemical abundance pattern of the brightest RGB member
found by \citet{kirby13} with a high-resolution Magellan/MIKE
spectrum.  In addition to the dwarf galaxy itself, \citet{belokurov09}
suggested that a higher metallicity stellar stream with a similar
velocity to Segue~2 is present in the same area of the sky.  With
their smaller survey area, \citet{kirby13} did not find evidence for
this additional population.  By combining the available data sets, we
find a total of 11 spectroscopically confirmed Segue~2 members and 280
confirmed non-members brighter than $g=20$.\footnote{One star,
  SDSSJ02192112+2007402, is classified as a non-member by
  \citet{belokurov09} and a blue horizontal branch member by
  \citet{kirby13}.  The magnitudes listed for this star by
  \citet{kirby13} do not match the cataloged magnitudes in DR7 or any
  subsequent data release of SDSS, suggesting an error or
  misidentification.  Since the SDSS magnitudes place the star well
  away from the Segue~2 horizontal branch and the \citet{belokurov09}
  velocity is not consistent with it being a Segue~2 member, I
  consider it a non-member here.}  In addition, 10 stars in the
possible foreground stream meet this magnitude cut.

\section{Hydrus~I}

The most recently discovered Milky Way satellite is Hydrus~I (Hyi~I),
announced just one week ago.  In the discovery paper,
\citet{koposov18} presented Magellan/M2FS spectroscopy of a sample of
$\sim30$ Hyi~I member stars.  While nearly all of these stars are
bright enough for \emph{Gaia} astrometry, the currently-available
preprint only lists the coordinates for 10 stars, the majority of
which are non-members and/or are fainter than $g=20$.
\citet{koposov18} also identified two RR~Lyrae variables in the
OGLE-IV Magellanic Cloud RR~Lyrae catalog \citep{soszynski16} as Hyi~I
members.  Combining the spectroscopic and RR~Lyrae lists, I have 4
confirmed Hyi~I members and 5 confirmed non-member stars brighter than
$g=20$.

\section{Horologium I}

Almost immediately after its discovery, \citet{koposov15b} published
VLT/FLAMES spectroscopy of Horologium~I (Hor~I).  \citet{nagasaw18}
used VLT/UVES to analyze the chemical abundance patterns of two of the
member stars identified by \citealt{koposov15b} as well as a third
star selected photometrically (which they confirmed as a Hor~I
member).  Combined, these two samples include 6
spectroscopically-confirmed Hor~I members and 13 confirmed non-members
brighter than $g=20$.

\section{Reticulum II}

\citet{simon15} observed Reticulum~II (Ret~II) with Magellan/M2FS and
Gemini-S/GMOS, which they combined with public VLT/FLAMES
spectroscopy from the \emph{Gaia}-ESO survey.  The target lists for the three
instruments largely overlapped; the VLT data include 6 stars (2
members) not observed by M2FS, and the GMOS data include 1 additional
member.  \citet{koposov15b} independently analyzed the VLT/FLAMES
spectra with essentially identical results.  A number of authors have
studied the chemical abundances of some of the many bright Ret~II
members with Magellan/MIKE or Magellan/M2FS
\citep{ji16,ji16c,roederer16,jf18}.  In total, there are 22 known
member stars and 23 non-members in Ret~II at $g < 20$.

\section{Carina II}

\citet{li18} medium-resolution spectroscopy of Carina~II (Car~II) with
Magellan/IMACS, AAT/AAOmega, and VLT/FLAMES and found 18 member stars.
Time-series imaging by \citet{torrealba18} identified 5 RR~Lyrae
variables in the field, of which 3 are at the correct distance to be
members of Car~II.  Two of the Car~II RR~Lyrae were spectroscopically
confirmed as members by \citet{li18}, and I assume that the third
star is a member as well.  Since no public photometry catalogs are
available for this part of the sky I use the dereddened magnitudes
given by \citet{li18} and remove the reddening correction to obtain
apparent magnitudes.  The 19 known Car~II members are all brighter
than $g=20$, as are a total of 8 confirmed non-members in the field.

\section{Carina III}

Carina~III (Car~III) was observed simultaneously with Car~II by
\citet{li18}, with 4 member stars identified.  The stars that are
spectroscopically determined to be members of neither Car~II nor
Car~III are mostly located closer to Car~II, but in order to provide
an astrometric comparison sample of non-members for Car~III I
consider the same set of 8 stars to be Car~III non-members.  As with
Car~II, I remove the reddening correction used by \citet{li18} to
determine the observed magnitudes of the Car~III stars.  Car~III has 3
confirmed member stars brighter than $g=20$.

\section{Ursa Major II}

\citet{martin07} and \citet[][hereafter SG07]{sg07} both observed
Ursa~Major~II (UMa~II) with Keck/DEIMOS.  Because of the large angular
size of the galaxy there is not much overlap between the two studies,
and combined they identified a total of 29 member stars.
\citet{frebel10} followed up the four brightest members on the RGB at
high-resolution with Keck/HIRES, finding that one of the four,
SDSSJ08500184+6313330, is actually a foreground main sequence star.
Time-series imaging by \citet{dallora12} uncovered one RR~Lyrae
variable in UMa~II, which is relatively bright (mean V magnitude of
18.39) given the distance of only 35~kpc to UMa~II.  UMa~II has 8
known members and 52 confirmed non-members brighter than $g=20$.

\section{Segue 1}

Segue~1 was spectroscopically confirmed by \citet{geha09}, who
identified 24 member stars with Keck/DEIMOS.  \citet{simon11} subsumed
the \citet{geha09} data set into a much larger DEIMOS spectroscopic
survey, finding a total of 71 Segue~1 members.  In the interim,
\citet{norris10} also obtained AAT spectroscopy of bright stars over a
wider field, including one additional Segue~1 member star outside the
\citet{simon11} coverage area and another member candidate at a radius
of 24\arcmin.  The entire sample of confirmed Segue~1 red giants has
been followed up at high-resolution by \citet{norris10c} and
\citet{fsk14}.  In total, Segue~1 contains 9 member stars brighter
than $g=20$ (7 RGB stars and 2 HB stars) plus the \citet{norris10}
candidate located at a radius of $5.5 r_{1/2}$.  Thanks to the
substantial spectroscopic investment in this field, there are 192
confirmed non-members at $g < 20$ as well.

\section{Ursa Major I}

Ursa~Major~I (UMa~I) has been the subject of three spectroscopic
studies.  Immediately after its discovery, \citet{kleyna05} obtained
low S/N, high-resolution spectroscopy of seven stars with the HIRES
spectrograph at Keck, five of which they determined were UMa~I
members.  \citet{martin07} observed a larger sample of stars at medium
resolution with Keck/DEIMOS, including two of the
\citeauthor{kleyna05} members.  Finally, SG07 observed UMa~I again
with DEIMOS and more than doubled the sample of member stars, as well
as re-observing all of the stars targeted by \citeauthor{kleyna05} and
most of the members identified by \citeauthor{martin07}.
All of the known bright ($g < 20$) members of UMa~I are in the SG07
data set, so I use that sample as my starting point for member
selection.  In total, there are eight spectroscopically confirmed
UMa~I members and 33 confirmed non-members at $g < 20$.

\section{Willman 1}

The first spectroscopy of Willman~1 was obtained by \citet{martin07}
with Keck/DEIMOS.  They classified 14 stars as Willman~1 members,
including two red giants brighter than $g=20$.  However,
high-resolution spectroscopy of one of the putative bright members
with the Hobby-Eberly Telescope by \citet{siegel08} showed that it is
instead a foreground main sequence star.  \citeauthor{siegel08}
confirmed the second RGB star as a member and obtained spectra of 3
non-members at large radii.  \citet{willman11} observed a much larger
sample with Keck/DEIMOS, including both candidates from the
\citet{martin07} study, and discovered an additional RGB member and
two horizontal branch stars.  Willman~1 contains four known members
and 22 confirmed non-members at $g<20$.

\section{Coma Berenices}

Coma Berenices (Com~Ber) has been observed spectroscopically with
Keck/DEIMOS by SG07, who found 59 member stars, and \citet{brown14},
who identified an additional 14 members.  \citet{frebel10} obtained
high-resolution spectroscopy of the brightest three SG07 members.  Two
RR~Lyrae variables in Com~Ber were discovered by \citet{musella09},
both of which are included in the combined spectroscopic sample from
SG07 and \citet{brown14}.  There are 12 known stars in Com~Ber
brighter than $g=20$ and 29 spectroscopically confirmed non-members
brighter than that limit.

\section{Bo{\"o}tes II}

The only multi-object spectroscopic study of Bo{\"o}tes~II (Boo~II)
published so far is that of \citet{koch09}, who identified a handful
of member stars with Gemini/GMOS.  \citet{kr14} followed up the
brightest \citeauthor{koch09} member star with Keck/HIRES, and
\citet{ji16b} obtained high-resolution Magellan/MIKE spectra of three
\citeauthor{koch09} members (including the star analyzed by
\citealt{kr14}) and a fourth bright member identified via
yet-unpublished spectroscopy.  The chemical abundances of two of these
member stars were also studied by \citet{francois16} with
VLT/X-Shooter spectra.  Despite being one of the least well
characterized ultra-faint dwarfs at present, Boo~II still contains 4
confirmed member stars and 4 confirmed non-members brighter than
$g=20$.

\section{Bo{\"o}tes I}
\label{sec:boo1mem}

Bo{\"o}tes I (Boo~I) has been the subject of numerous spectroscopic
studies at both medium and high spectral resolution.  Radial
velocities have been measured by \citet[][WIYN/Hydra]{munoz06},
\citet[][Keck/DEIMOS]{martin07}, \citet[][AAT/AAOmega]{norris10},
\citet[][VLT/FLAMES]{koposov11}, and \citet[][Keck/DEIMOS]{brown14}.
The chemical abundances of Boo~I stars have been studied by
\citet{feltzing09}, \citet{norris08,norris10,norris10b},
\citet{lai11}, \citet{gilmore13}, \citet{ishigaki14}, and
\citet{frebel16}.  Time-series imaging of Boo~I has also identified 15
RR~Lyrae variables in the galaxy \citep{siegel06,dallora06}.  As a
result of its relatively high luminosity and the extensive follow-up
data sets available for Boo~I, there are 68 confirmed member stars and
138 confirmed non-members at $g < 20$.

\section{Draco II}

Draco~II (Dra~II) ranks with Boo~II as one of the most poorly-studied
of the nearby ultra-faint dwarfs.  \citet{martin16b} obtained
Keck/DEIMOS spectroscopy of Dra~II, identifying 9 member stars and 25
non-members.  Only four of the members are brighter than $g=20$, with
the brightest at $g=19.4$.  No high-resolution chemical abundance
studies are available.  The online table in \citet{martin16b}
containing the list of spectroscopic non-members is not currently
available, so our sample for the present work includes only the 4
bright members.

\section{Tucana II}

Tucana~II (Tuc~II) was observed spectroscopically with Magellan/M2FS
by \citet{walker16}, who identified 8 member stars.  Four of these
stars were followed up at higher spectral resolution with
Magellan/MIKE to determine their chemical abundances by \citet{ji16d}.
\citet{chiti18} observed one additional \citeauthor{walker16} member,
as well as two new member stars selected with SkyMapper photometry,
with Magellan/MIKE, so that chemical abundance patterns are now
available for the majority of the known Tuc~II members.  I adopt a
membership probability threshold of $p \ge 0.5$ to separate members
and non-members in the \citeauthor{walker16} catalog, but the
resulting classifications are not very sensitive to this value (e.g.,
$p \ge 0.1$ produces identical lists of stars).  The union of these
data sets results in a total of 10 member stars and 49 confirmed
non-members brighter than $g=20$.

\section{Tucana III}

\citet{simon17} obtained medium-resolution Magellan/IMACS spectroscopy
of Tucana~III (Tuc~III), identifying 26 member stars within the core
of the satellite.  \citet{hansen17} studied the chemical abundances of
the brightest Tuc~III star with a high-resolution Magellan/MIKE
spectrum.  Ten out of the 26 spectroscopically-confirmed members are
brighter than $g=20$, and \citet{simon17} also classified 53 stars at
$g<20$ as spectroscopic non-members.

\section{Hydrus~I Astrometric Member Selection}
\label{hyi1_select}

\begin{figure*}[th!]
\epsscale{1.15}
\plottwo{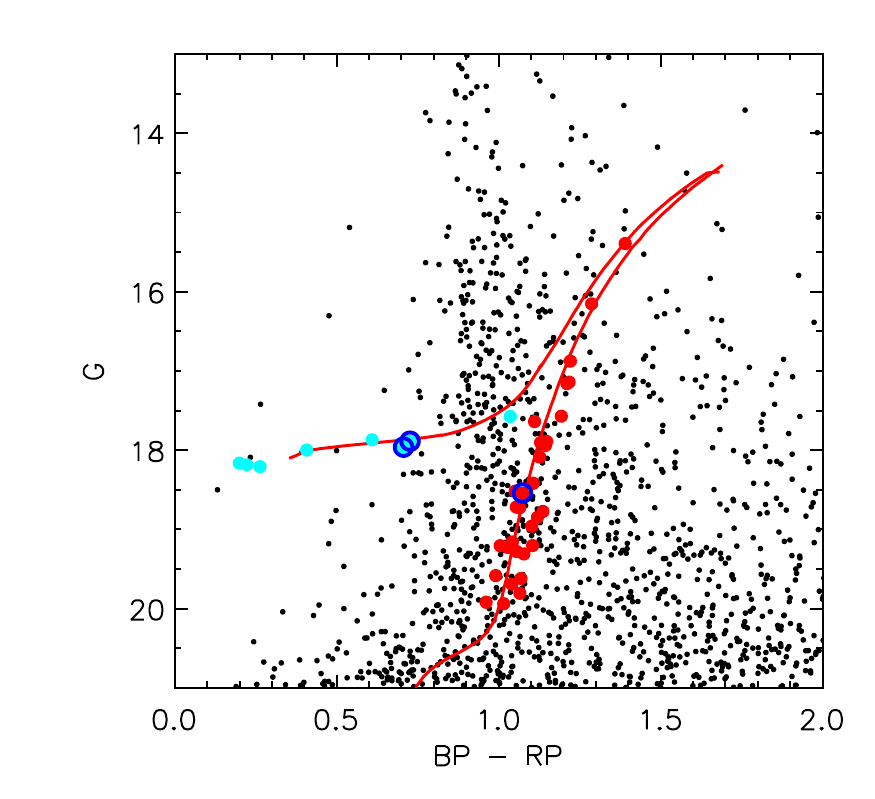}{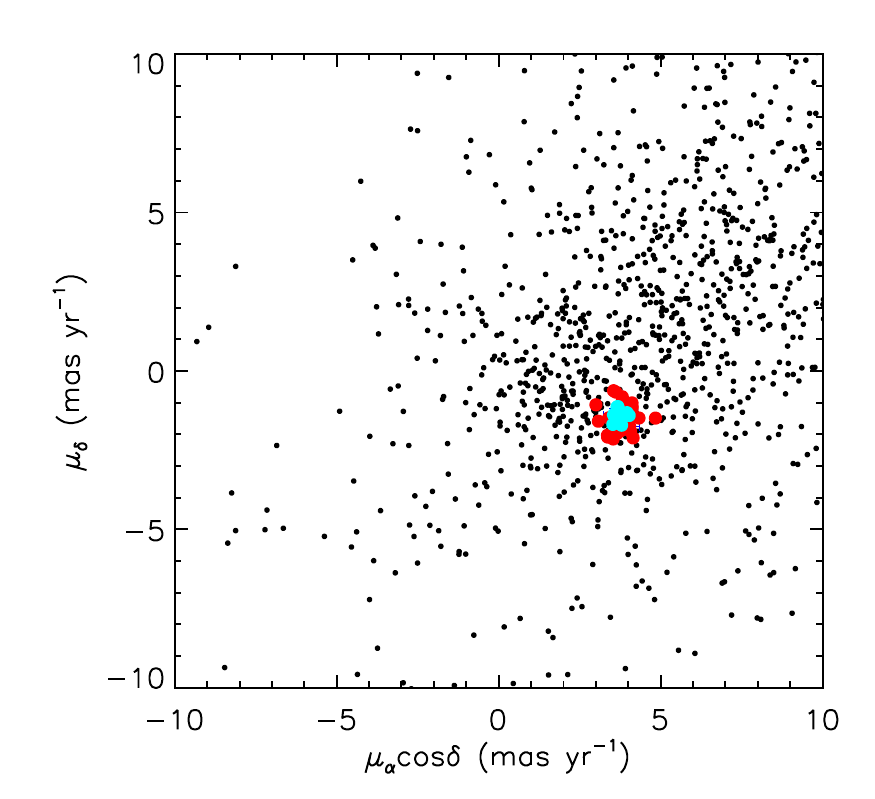}
\caption{(left) \emph{Gaia} color-magnitude diagram of Hyi~I.  The 3
  known member stars are outlined in blue, and the proper motion +
  color-selected candidates are displayed as filled red (cyan) circles
  on the RGB (HB).  The red curve is a PARSEC isochrone an age of
  12~Gyr and a metallicity of $\feh = -2.0$.  (right)
  Proper motion diagram of the same stars.}
\label{hyi1plot}
\end{figure*}

As discussed earlier, the sample of members stars currently available
in Hyi~I is only 3.  However, using the proper motion of Hyi~I
determined from these 3 stars, a likely complete member sample of
Hyi~I can be selected from the \emph{Gaia} DR2 catalog.  In the left
panel of Figure~\ref{hyi1plot} I plot the \emph{Gaia} color-magnitude
diagram (CMD) of a field with radius 15\arcmin\ centered on Hyi~I.
The three stars outlined with blue circles are the members identified
previously.  The red curve is a PARSEC isochrone \citep{marigo17} for
a 12~Gyr old population at $\feh = -2.0$, shifted to a distance
modulus of 17.20.  Because the reddening for faint stars in the
\emph{Gaia} bands is not yet well-determined, I assume $A_{G} = 2
\times E(BP-RP)$ \citep{andrae18} and estimate the extinction
empirically by finding the values for which the isochrone matches both
the horizontal branch and RGB stars ($A_{G} = 0.36$~mag, $E(BP-RP) =
0.18$~mag).  Choosing the stars with proper motions within
1~\masyr\ of the average value for the \citet{koposov18} members
produces a clear set of red giant branch stars in the CMD.  As a final
step, I select the stars that are both within 0.08~mag of the RGB
(0.30~mag on the HB) and have proper motions within $2\sigma$ of the
Hyi~I value in Table~\ref{pmdata_table}.  I find 41 such stars
brighter than $G=20$ in the DR2 catalog, which trace out a
well-defined RGB and horizontal branch in the CMD, and are tightly
clustered in proper motion space (Fig.~\ref{hyi1plot}).  In this way,
starting from just 3 member stars one can identify a highly pure
sample of member stars that will greatly improve the efficiency of
future spectroscopic observations.  Of course, Hyi~I is a particularly
easy case given its relatively high luminosity and modest distance,
but similar methods can certainly be applied to other dwarfs.  With
this 41 star sample the proper motion uncertainties for Hyi~I are
reduced by a factor of $\sim5$ to $\sim0.03$~\masyr\ in each
coordinate, close to the DR2 systematic floor.  However, the orbit is
essentially unchanged.

\end{appendix}
\clearpage

\end{document}